\documentclass{sig-alternate}
\usepackage{amssymb}
\usepackage{xspace}
\usepackage{graphicx}
\usepackage{psfrag}
\usepackage{url}

\usepackage{hyperref}
\usepackage{breakurl}
\usepackage{pifont}
\usepackage{subfigure}
\usepackage{adjustbox}
\usepackage{epstopdf}
\usepackage{algpseudocode}
\usepackage{algorithm}
\usepackage{comment}
\usepackage{enumitem}
\usepackage[english]{babel}
\usepackage{mathptmx}
\usepackage{tablefootnote} \usepackage[usenames,dvipsnames]{xcolor}
\usepackage{array} \usepackage{balance}

\DeclareSymbolFont{operators}{OT1}{cmr}{m}{n}
\DeclareSymbolFont{letters}{OML}{cmm}{m}{it}
\DeclareSymbolFont{symbols}{OMS}{cmsy}{m}{n}
\DeclareSymbolFont{largesymbols}{OMX}{cmex}{m}{n}

\def\centerhack#1{\hbox to 0pt{\hss\footnotesize #1\hss}}

\renewcommand{\baselinestretch}{0.99}

\makeatletter
\def\@listi{\leftmargin\leftmargini
    \parsep 1\p@ \@plus0\p@ \@minus\p@
    \topsep 2\p@   \@plus0\p@ \@minus\p@
    \itemsep1\p@ \@plus0\p@ \@minus\p@}
\let\@listI\@listi\@listi
\makeatother

\newcommand{\etal}{et~al.\@\xspace} 
\newcommand{\eg}{e.g.,\xspace}
\newcommand{\ie}{i.e.,\xspace}

\newcommand{\name}{HORNET\xspace}

\newcommand{\lap}{LAP\xspace}
\newcommand{\dovetail}{Dovetail\xspace}
\newcommand{\sphinx}{Sphinx\xspace}
\newcommand{\scion}{SCION\xspace}

\newcommand{\nira}{NIRA\xspace}
\renewcommand\paragraph[1]{\smallskip\noindent\textbf{#1.}}
\newcommand{\setupfirst}{\textsf{P}\ding{202}\xspace}
\newcommand{\setupsecond}{\textsf{P}\ding{203}\xspace}

\newcommand{\MAC}{\ensuremath{\mathsf{MAC}}}
\newcommand{\PRG}{\ensuremath{\mathsf{PRG0}}}
\newcommand{\PRP}{\ensuremath{\mathsf{PRP}}}
\newcommand{\PRGTWO}{\ensuremath{\mathsf{PRG1}}}
\newcommand{\PRGTHREE}{\ensuremath{\mathsf{PRG2}}}
\newcommand{\ENC}{\ensuremath{\mathsf{ENC}}}
\newcommand{\DEC}{\ensuremath{\mathsf{DEC}}}
\newcommand{\randomgen}{{\sc rand}}

\newcommand{\aheader}{\textsc{ahdr}\xspace}

\newcommand{\tor}{Tor\xspace}

\newcommand{\FS}{\ensuremath{\mathit{FS}}\xspace}

\newcommand{\concat}{\ensuremath{\mathbin{\|}}}
\makeatletter
\newcommand{\StatexIndent}[1][3]{	\setlength\@tempdima{\algorithmicindent}	\Statex\hskip\dimexpr#1\@tempdima\relax}
\makeatother

\def\centerhack#1{\hbox to 0pt{\hss\footnotesize #1\hss}}
\def\dchack#1{\vbox to 0pt{\vss{\hbox to 0pt{\hss#1\hss}}\vss}}

\usepackage{float}
\newfloat{algorithm}{t}{lop}

\title{\name: High-speed Onion Routing at the Network Layer}
\author{
	Chen Chen\\
	chen.chen@inf.ethz.ch\\
	CMU/ETH Z\"urich
\and
	Daniele E. Asoni\\
	daniele.asoni@inf.ethz.ch\\
	ETH Z\"urich
\and
	David Barrera\\
	david.barrera@inf.ethz.ch\\
	ETH Z\"urich
\and
	George Danezis\\
	g.danezis@ucl.ac.uk\\
	University College London
\and
	Adrian Perrig\\
	adrian.perrig@inf.ethz.ch\\
	ETH Z\"urich
}

\newfont{\mycrnotice}{ptmr8t at 7pt}
\newfont{\myconfname}{ptmri8t at 7pt}
\let\crnotice\mycrnotice\let\confname\myconfname
\permission{Permission to make digital or hard copies of all or part of this work for personal or classroom use is granted without fee provided that copies are not made or distributed for profit or commercial advantage and that copies bear this notice and the full citation on the first page. Copyrights for components of this work owned by others than the author(s) must be honored. Abstracting with credit is permitted. To copy otherwise, or republish, to post on servers or to redistribute to lists, requires prior specific permission and/or a fee. Request permissions from Permissions@acm.org.}
\conferenceinfo{CCS'15,}{October 12--16, 2015, Denver, Colorado, USA. \\ 
	{\mycrnotice{Copyright is held by the owner/author(s). Publication rights licensed to ACM.}}}
\copyrightetc{ACM \the\acmcopyr}
\crdata{978-1-4503-3832-5/15/10\ ...\$15.00.\\
	DOI: http://dx.doi.org/10.1145/2810103.2813628}

\makeatletter
\def\@copyrightspace{\relax}
\makeatother

\makeatletter
\def\thebibliography#1{	\ifnum\addauflag=0\addauthorsection\global\addauflag=1\fi
	\section[References]{		{References} 		{\vskip -2pt plus 1pt} 		\@mkboth{{\refname}}{{\refname}}	}	\list{[\arabic{enumi}]}{		\settowidth\labelwidth{[#1]}		\leftmargin\labelwidth
		\advance\leftmargin\labelsep
		\advance\leftmargin\bibindent
		\parsep=0pt\itemsep=1pt 		\itemindent -\bibindent
		\listparindent \itemindent
		\usecounter{enumi}
	}	\let\newblock\@empty
	\raggedright 	\sloppy
	\sfcode`\.=1000\relax
}
\makeatother

\begin{document}

\maketitle

\begin{abstract}
We present HORNET, a system that enables high-speed end-to-end anonymous
channels by leveraging next-generation network architectures. HORNET is
designed as a low-latency onion routing system that operates at the network
layer thus enabling a wide range of applications. Our system uses only
symmetric cryptography for data forwarding yet requires no per-flow state on
intermediate routers. This design enables HORNET routers implemented on
off-the-shelf hardware to process anonymous traffic at over 93~Gb/s. \name is
also highly scalable, adding minimal processing overhead per additional
anonymous channel. 

\end{abstract}




\section{Introduction}
\label{sec:introduction}
\begin{sloppypar}
Recent revelations about global-scale pervasive surveillance~\cite{pervasive_monitoring_rfc} programs have 
demonstrated that the privacy of Internet users worldwide is at risk. These 
revelations suggest massive amounts of private traffic, including web browsing 
activities, location information, and personal communications are being 
harvested in bulk by domestic and foreign intelligence agencies. 

To protect against these threats, several anonymity 
protocols, tools, and architectures have been proposed. Among the most secure 
schemes for anonymous communications are mix 
networks~\cite{DBLP:conf/ndss/GulcuT96,mixmaster-spec,DBLP:conf/sp/DanezisDM03,Danezis2009},
 which provide high-latency asynchronous messaging.
 Onion routing networks (most notably Tor~\cite{dms04}), offer a 
balance between security and performance, enabling low-latency anonymous 
communication suitable for typical Internet activities (\eg web browsing, 
instant messaging, etc.). 
Tor is the system of choice for over 2 million daily 
users~\cite{tormetric}, but its design as an overlay 
network suffers from performance and scalability issues.
Tor's design 
requires per-connection state to be maintained by intermediate nodes, 
limiting the total number of concurrent anonymous connections that can take 
place simultaneously.  
\end{sloppypar}

The scalability and performance limitations of anonymous networks have been
partially addressed by building protocols into the network layer rather than
implementing them as overlays. Among these high-performing schemes are
LAP~\cite{Hsiao2012} and Dovetail~\cite{Sankey2014}, which offer network-level
low-latency anonymous communication on next-generation network architectures.
The high performance of both schemes, however, results in significantly
degraded security guarantees; endpoints have little to no protection against
adversaries that are not confined to a single network location, and payload
protection relies on upper layer protocols which increases complexity.

In this paper, we present \name (High-speed Onion Routing at the NETwork
layer), a highly-scalable anonymity system that leverages next-generation
Internet architecture design. \name offers payload protection by default, and
can defend against attacks that exploit multiple network observation points.
\name is designed to be highly efficient:
it can use short paths offered by underlying network architectures, rather than the long paths due to global redirection; additionally, instead of keeping state at each
relay, connection state (including, \eg onion layer decryption keys) is carried
within packet headers, allowing intermediate nodes to quickly forward traffic
without per-packet state lookup.

While this paper proposes and evaluates a concrete anonymity system, a 
secondary goal herein is to broadly re-think the design of low-latency 
anonymity systems by envisioning networks where anonymous communication is 
offered as an in-network service to all users.
For example, what performance trade-offs exist between keeping anonymous
connection state at relays and carrying state in packets? If routers  perform
anonymity-specific tasks, how can we ensure that these operations do not impact
the processing of regular network traffic, especially in adversarial
circumstances? And if the network architecture should provide some support for
anonymous communication, what should that support be? Throughout the paper we
consider these issues in the design of our own system, and provide intuition
for the requirements of alternative network-level anonymity systems. 

Specifically, our contributions are the following:
\begin{itemize}

\item We design and implement \name, an anonymity system  that uses 
source-selected paths and shared keys between endpoints and routers to support
onion routing. Unlike other onion routing implementations, \name routers do not 
keep per-flow state or perform computationally expensive operations for data 
forwarding, allowing the system to scale.

\item We analyze the security of \name, showing that it can defend against 
passive attacks, 
and certain types of active attacks. \name provides 
stronger security guarantees than existing network-level anonymity systems. 

\item We evaluate the performance of \name, showing that its anonymous data
    processing speed is close to that of LAP and Dovetail (up to 93.5 Gb/s
    on a 120 Gb/s software router).
    This performance is comparable with that of today's high-end commodity routers~\cite{ciscorouters}.

\end{itemize}

\section{Problem Definition}
\label{sec:probldef}

We aim to design a network-level anonymity system to frustrate adversaries with
mass surveillance capabilities. Specifically, an adversary observing traffic
traversing the network should be unable to link (at large scale) pairs of
communicating hosts . This property is known as relationship
anonymity~\cite{pfitzmann2001}.

We define \emph{sender anonymity} as a communication scenario where anonymity is guaranteed for the source, but the destination's location is 
public (\eg web sites for The Guardian or Der Spiegel). We define 
\emph{sender-receiver anonymity} as a scenario where the anonymity guarantee is extended to the destination (\eg a hidden service that wishes to conceal its location). Sender-receiver anonymity therefore offers protection for both ends, implying sender anonymity.
Depending on users' needs, \name can support either sender anonymity or sender-receiver anonymity.

Since our scheme operates at the network layer, network location is the only 
identity feature we aim to conceal. Exposure of network location or user 
identity at upper layers (\eg through TCP sessions, login credentials, or 
browser cookies) is out of scope for this work.

\subsection{Network Model}
\label{sec:network_model}
We consider that provisioning anonymous communication between end users 
is a principal task of the network infrastructure. The network's anonymity-related infrastructures,
primarily routers, assist end users in establishing temporary \emph{anonymous sessions} for anonymous data transmission.

We assume that the network layer is operated by a set of nodes.  Each node
cooperates with sources to establish anonymous sessions to the intended
destinations, and processes anonymous traffic within the created sessions. We
require that the routing state of a node allows it to determine only the next
hop. In particular, the destination is only revealed to the last node and no
others. This property can be satisfied by IP Segment
Routing~\cite{segment_routing}, Future Internet Architectures (FIAs) like
NIRA~\cite{Yang2007NIRA} and SCION~\cite{Xin2011SCION,scion2015}, or
Pathlets~\cite{godfrey2009pathlet}.  In practice, our abstract notion of a node
could correspond to different entities depending on the architecture on which
\name is built. For instance, in \nira and \scion, a node corresponds to an
Autonomous System (AS); in Pathlets, a node maps to a \emph{vnode}.

\paragraph{Path and certificate retrieval}
A path is the combination of routing state of all nodes between the source and
the intended destination. We assume the underlying network architecture
provides a mechanism for a source to obtain such a path to a given destination.
Additionally, we assume that the same mechanism allows the source to fetch the
public keys and certificates\footnote{Depending on the underlying PKI scheme,
the source might need to fetch a chain of certificates leading to a trust
anchor to verify each node's public key.} of on-path nodes.  Note that the
mechanism should be privacy-preserving: the source should not reveal its
network location or intent to communicate with a destination by retrieving
paths, public keys, and certificates. In
Section~\ref{sec:discussion:retrieve_path}, we further discuss how to obtain
required information anonymously in selected FIAs. While a general solution
represents an important avenue for future work, it remains outside of our
present scope.

\paragraph{Public key verification}
We assume that end hosts and on-path nodes have public keys accessible and verifiable by all entities. End hosts can retrieve the public keys of other end hosts through an out-of-band channel (e.g., websites) and verify them following a scheme like HIP~\cite{hip}, in which the end hosts can publish hashes of their public keys as their service names. Public keys of on-path nodes are managed through a public-key infrastructure (PKI). For example, the source node can leverage Resource Public Key Infrastructure (RPKI)~\cite{rpki} to verify the public keys of on-path nodes.

\subsection{Threat Model}

We consider an adversary attempting to conduct mass surveillance. Specifically,
the adversary collects and maintains a list of ``selectors'' (\eg targets'
network locations, or higher-level protocol identifiers), which help the
adversary trawl intercepted traffic and extract parts of it for more extensive
targeted analysis~\cite{selector_story}.
An anonymity system should prevent an adversary from leveraging
bulk communication access to select traffic that belongs to the targets.
Thus an adversary has to collect and analyze all traffic and cannot
reliably select traffic specific to targets unless it has access to
the physical links adjacent to the targets.

We consider an adversary that is able to compromise a fraction
of nodes on the path between
a source and a destination. For sender anonymity, the adversary can also compromise
the destination. For sender-receiver anonymity, the adversary can compromise at
most one of the two end hosts.
By compromising a node, the adversary learns all keys and settings, observes all traffic that 
traverses the compromised node, and is able to control how the nodes behave including
redirecting traffic, fabricating, replaying, and modifying packets. 

However, we do not aim to prevent targeted de-anonymization attacks where an adversary invests a significant amount of resources on a single or a small set of victims.
Like other low-latency schemes, we cannot solve targeted confirmation attacks based on the analysis of flow dynamics~\cite{SS03, timing-fc2004, murdoch-pet2007}.
Defending against such attacks using dynamic link padding~\cite{wang2008dependent} would be no more difficult than in onion routing, although equally expensive. We defer the discussion and analysis of such measures to future work.

\subsection{Desired Properties}
\label{sec:desired_properties}
\name is designed to achieve the following anonymity and security properties:
\begin{enumerate}
	\item \textbf{Path information integrity and secrecy}. 
	An adversary cannot modify a packet header to alter a network path without detection.
	The adversary should not learn forwarding information of uncompromised nodes, node's positions, or the total number of hops on a path. 	
    \item \textbf{No packet correlation.} An adversary who can eavesdrop on
        multiple links in the network cannot correlate packets on those links
        by observing the bit patterns in the headers or payloads. This should
        hold regardless of whether the observed traffic corresponds to the same
        packet (at different points on the network), or corresponds to different packets from a single session. 
        \item \textbf{No session linkage.} 
		An adversary cannot link packets from different sessions, even between
    the same source and destination. 
	\item \textbf{Payload secrecy and end-to-end integrity}. Without compromising end hosts, 
	an adversary cannot learn any information from the data payload except for its length and timing among sequences of packets. 
	
	\end{enumerate}

\section{\name Overview}
\label{sec:overview}
 \newcommand{\cmnHdr}{\mbox{\sc chdr}}
\newcommand{\CmnHdr}{$\cmnHdr$\xspace}
\newcommand{\expTime}{\mbox{\sc exp}}
\newcommand{\ExpTime}{$\expTime$\xspace}
The basic design objectives for \name are \emph{scalability} and \emph{efficiency}. To enable Internet-scale anonymous communication, \name intermediate nodes must avoid keeping per-session state (\eg cryptographic keys and routing information). Instead, session state is offloaded to end hosts, who then embed this state into packets such that each intermediate node can extract its own state as part of the packet forwarding process.

Offloading the per-session state presents two challenges. First, nodes need to prevent their offloaded state from leaking information (\eg the session's cryptographic keys). To address this, each \name node maintains a local secret to encrypt the offloaded per-session state. We call this encrypted state a \emph{Forwarding Segment} (FS). The FS allows its creating node to dynamically retrieve the embedded
information (\ie next hop, shared key, session expiration time), while hiding this information from unauthorized third parties.

The second challenge in offloading the per-session state is to combine this state (\ie the FSes) in a packet in such a way that each node is able to retrieve its own FS, but no information is leaked about the network location of the end hosts, the path length, or a specific node's position on the path. Learning any of this information could assist in de-anonymization attacks (see Section~\ref{sec:ilt}).
To address this challenge, the source constructs an \emph{anonymous header} (\aheader) by combining multiple FSes, and prepends this header to each packet in the session. An \aheader grants each node on the path access to the FS it created, without divulging any information about the path except for a node's previous and next nodes (see Section~\ref{sec:aheader}).

For efficient packet processing, each \name node performs one Diffie-Hellman (DH) key exchange operation once per session during setup.
For all data packets within the session, \name nodes use only symmetric cryptography to retrieve their state, process the \aheader and onion-decrypt (or encrypt) the payload.  
To reduce setup delay, \name uses only two setup packets within a single round trip between the source and the destination. Therefore, session setup only incurs $O(n)$ propagation delay in comparison to $O(n^2)$ by the telescopic setup method used in Tor (where $n$ is the number of anonymity nodes traversed on the path).
While for Tor the default value of $n$ is 3, for \name $n$ might be as large as 14 (4.1 in the average case, and less or equal to 7 in over 99\% of cases~\cite{iplane_dataset}),
which emphasizes the need to optimize setup propagation delay.

\subsection{Sender Anonymity}
\label{sec:senderanonymity}
Anonymous sessions between a source and a destination require the source to establish state between itself and every node on the path. The state will be carried in subsequent data packets, enabling intermediate nodes to retrieve their corresponding state and forward the packet to the next hop. We now describe how the state is collected without compromising the sender's anonymity, and how this state is used to forward data packets. 

\paragraph{Setup phase}
To establish an anonymous session between a source $S$ and a public destination $D$, $S$ uses a single round of Sphinx~\cite{Danezis2009}, a provably secure mix protocol (an overview of Sphinx is given in Section~\ref{sec:sphinx}). This round consists of two Sphinx packets (one for the forward path and one for the backward path) each of which will anonymously establish shared symmetric keys between $S$ and every node on that path. For \name, we extend the Sphinx protocol to additionally anonymously collect the forwarding segments (FSes) for each node. Our modified Sphinx protocol protects the secrecy and integrity of these FSes, and does not reveal topology information to any node on the path. We note that using Sphinx also for data forwarding would result in low throughput due to prohibitively expensive per-hop asymmetric cryptographic operations. Therefore, we use Sphinx only for session setup packets, which are amortized over the subsequent data transmission packets. We explain the details of the setup phase in Section~\ref{sec:setupphase}.

\paragraph{Data transmission phase}
Having collected the FSes, the source is now able to construct a forward
\aheader and a backward \aheader for the forward and backward paths,
respectively. \aheader{}s carry the FSes which contain all state necessary for
nodes to process and forward packets to the next hop. When sending a data
packet, the source onion-encrypts the data payload using the session's shared
symmetric keys, and prepends the \aheader. Each node then retrieves its FS from the \aheader, onion-decrypts the packet
and forwards it to the next hop, until it reaches the destination. The
destination uses the backward \aheader (received in the first data
packet\footnote{If the first packet is lost the source can simply resend the
backward \aheader using a new data packet (see
Section~\ref{sec:datatransmission}).}) to send data back to $S$, with the only
difference being that the payload is encrypted (rather than decrypted) at each
hop. We present the details of the data transmission phase in
Section~\ref{sec:datatransmission}.

\subsection{Sender-Receiver Anonymity}
\label{sec:recv_anonymity}
Sender-receiver anonymity, where neither $S$ nor $D$ knows the other's location
(\eg a hidden service), presents a new challenge: since $S$ does not know $D$'s location (and vice versa), $S$ cannot retrieve a path to $D$, precluding the establishment of state between $S$ and nodes on the path to $D$ as described in Section~\ref{sec:senderanonymity}. 

A common approach to this problem (as adopted by \tor\footnote{\tor
additionally uses an introduction point, which enables $S$ to negotiate a
 rendezvous point with $D$. 
This design provides additional scalability and attack
resistance~\cite{dms04}, but increases the delay of setting up a session.
\name's design favors simplicity and performance, but nothing fundamentally
prevents \name from using \tor's approach.}, \lap, and \dovetail) is to use a
public \emph{rendezvous point} (RP) to forward traffic between $S$ and $D$
without knowing either $S$ or $D$.  This solution would also work for \name, but
would require RPs to maintain per-session state between sources and
destinations. For instance, when receiving a packet from $S$, an RP needs the
state to determine how to send the packet to $D$.  Maintaining per-session
state on RPs increases complexity, bounds the number of receivers, and
introduces a state exhaustion denial-of-service attack vector. 

\noindent\textbf{Nested \aheader{}s.} Our proposal for sender-receiver anonymity requires no state to be kept at the RP by nesting the necessary state for RPs to forward a packet  within the packet's header: a forward \aheader from $S$ to a RP will include the \aheader from the RP to $D$; a backward \aheader from $D$ to a RP will include the \aheader from the RP back to $S$.  

Briefly, to establish a \name session between $S$ and $D$ keeping both parties
hidden from each other, $D$ selects a public rendezvous point $R$ and completes
a \name session setup between $D$ and $R$. $D$ publishes
$\aheader_{R\rightarrow{}D}$ to a public directory. Note that this \aheader
leaks no information about $D$'s location and can only be used to send data to
$D$ through $R$ within a specific time window.

When $S$ wants to send traffic to $D$, $S$ retrieves (from a public directory) $\aheader_{R\rightarrow{}D}$. $S$ then establishes a \name session between $S$ and $R$ and constructs a nested \aheader with $\aheader_{R\rightarrow{}D}$ inside $\aheader_{S\rightarrow{}R}$. Thus, when $R$ receives a packet from $S$, $R$ can retrieve $\aheader_{R\rightarrow{}D}$ from $\aheader_{S\rightarrow{}R}$ and forward the packet to $D$. 
$S$ also includes $\aheader_{R\rightarrow{}S}$ in the data payload of the first data packet to $D$, allowing $D$ to create a return path to $S$. 

One of the advantages of our scheme is that any node on the network can serve as a rendezvous point. In fact, multiple points can be selected and advertised, allowing the source to pick the RP closest to it. Moreover, once a \name session has been established, $S$ and $D$ can negotiate a better (closer) RP (\eg using private set intersection~\cite{Freedman2004}). A  disadvantage of the nested \aheader technique is that it doubles the size of the header.

\subsection{Packet Structure}
\label{sec:packetstructure}

\begin{figure}[!htbp]
	\centering
	\includegraphics[width=8cm]{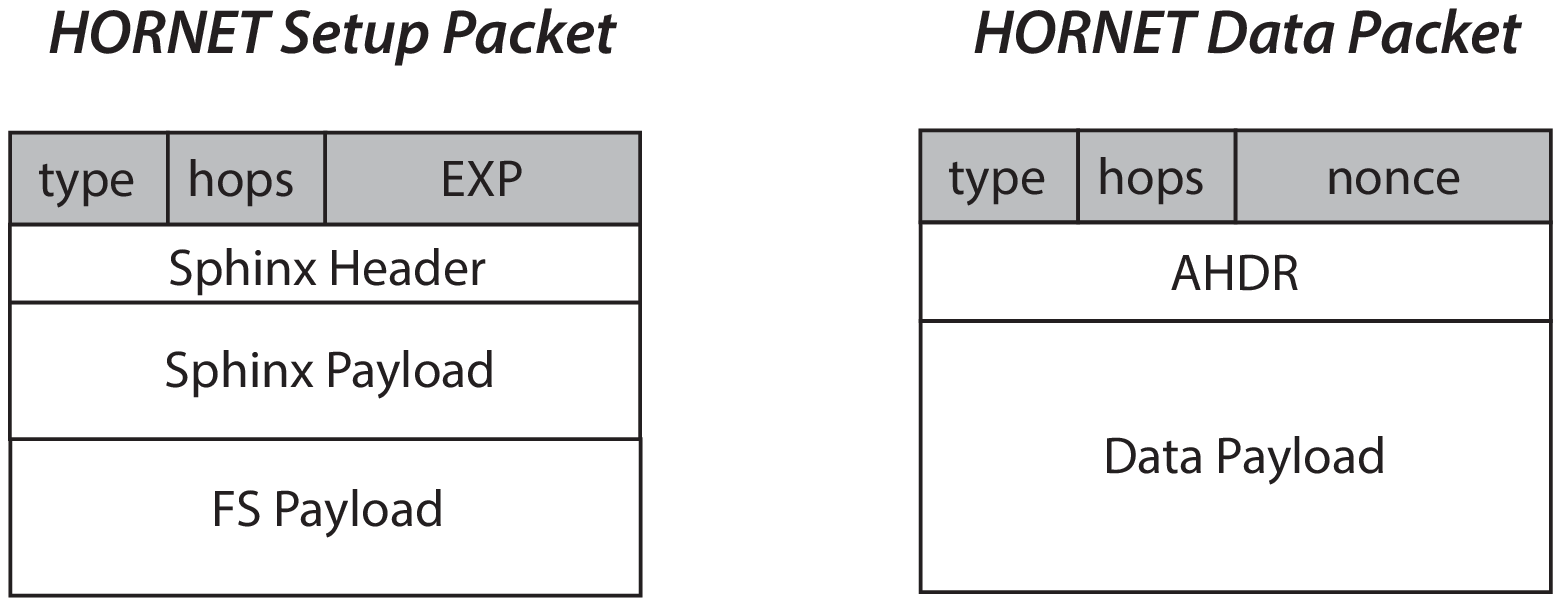}
	\caption{\name packet formats. For both setup packet and data packet, the shaded fields represent the common header (\CmnHdr).}
	\label{fig:setup_pkts}
\end{figure}

\name uses two types of packets: \emph{setup packets} and \emph{data packets} (see Figure~\ref{fig:setup_pkts}). 
Both types of packets begin with a common header (\CmnHdr)
which describes the packet type, the length of the longest path that the session supports, and a type-specific field. 
For session setup packets, the type-specific field contains a value $\expTime$ which indicates 
the intended expiration time of the session. For data packets, the specific value is a
random nonce generated by the sender used by intermediate nodes to process the data packet.

Session setup packets include a nested \sphinx packet and an FS payload. Data packets carry an \aheader and an onion-encrypted data payload. We explain each field in detail in Section~\ref{sec:protocol}.

\section{Formal Protocol Description}
\label{sec:protocol}
\newcommand{\hdr}{\mbox{\sc HDR}}
\newcommand{\sphinxHdr}{{\mbox{{\sc shdr}}}}
\newcommand{\SphinxHdr}{$\sphinxHdr$\xspace}
\newcommand{\sphinxHdrSend}{{\sphinxHdr^{f}}}
\newcommand{\SphinxHdrSend}{$\sphinxHdrSend$\xspace}
\newcommand{\sphinxHdrRecv}{{\sphinxHdr^{b}}}
\newcommand{\SphinxHdrRecv}{$\sphinxHdrRecv$\xspace}
\newcommand{\sphinxPayload}{{\mbox{\sc sp}}}
\newcommand{\SphinxPayload}{$\sphinxPayload$\xspace}

\newcommand{\SphinxPayloadSend}{$\sphinxPayload^{f}$\xspace}
\newcommand{\SphinxPayloadRecv}{$\sphinxPayload^{b}$\xspace}
\newcommand{\sphinxGenSphinxHdr}{\mbox{{\sc gen\_sphx\_hdr}}}
\newcommand{\SphinxGenSphinxHdr}{$\sphinxGenSphinxHdr$\xspace}
\newcommand{\sphinxGenSphinxPayloadSend}{\mbox{{\sc gen\_sphx\_pl\_send}}}
\newcommand{\SphinxGenSphinxPayloadSend}{$\sphinxGenSphinxPayloadSend$\xspace}
\newcommand{\sphinxUnwrapSphinxPayloadSend}{\mbox{{\sc unwrap\_sphx\_pl\_send}}}
\newcommand{\SphinxUnwrapSphinxPayloadSend}{$\sphinxUnwrapSphinxPayloadSend$\xspace}
\newcommand{\sphinxUnwrapSphinxPayloadRecv}{\mbox{{\sc unwrap\_sphx\_pl\_recv}}}
\newcommand{\SphinxUnwrapSphinxPayloadRecv}{$\sphinxUnwrapSphinxPayloadRecv$\xspace}
\newcommand{\sphinxGenSphinxPayloadRecv}{\mbox{{\sc gen\_sphx\_pl\_recv}}}
\newcommand{\SphinxGenSphinxPayloadRecv}{$\sphinxGenSphinxPayloadRecv$\xspace}
\newcommand{\sphinxProcSphinxPkt}{\mbox{{\sc proc\_sphx\_pkt}}}
\newcommand{\SphinxProcSphinxPkt}{$\sphinxProcSphinxPkt$\xspace}

\newcommand{\sphinxPerNodePubKey}{{x_i}}
\newcommand{\sphinxPerNodePrivKey}{{y_i}}

\newcommand{\fsCreate}{\mbox{{\sc fs\_create}}}
\newcommand{\FsCreate}{$\fsCreate$\xspace}
\newcommand{\fsOpen}{\mbox{{\sc fs\_open}}}
\newcommand{\FsOpen}{$\fsOpen$\xspace}

\newcommand{\fsPayloadInit}{\mbox{{\sc init\_fs\_payload}}}
\newcommand{\addFs}{\mbox{\sc add\_fs}}
\newcommand{\retrieveFs}{\mbox{\sc retrieve\_fses}}
\newcommand{\FsPayloadInit}{$\fsPayloadInit$\xspace}
\newcommand{\AddFs}{$\addFs$\xspace}
\newcommand{\RetrieveFs}{$\retrieveFs$\xspace}
\newcommand{\fsPayload}{{P}}
\newcommand{\FsPayload}{$\fsPayload$\xspace}
\newcommand{\fsPayloadRecv}{\fsPayload^{b}}
\newcommand{\FsPayloadRecv}{$\fsPayloadRecv$}

\renewcommand{\cmnHdr}{\mbox{\sc chdr}}
\renewcommand{\CmnHdr}{$\cmnHdr$\xspace}
\providecommand{\expTime}{\mbox{\sc exp}}
\providecommand{\ExpTime}{$\expTime$\xspace}

\newcommand{\aHdr}{\mbox{{\sc ahdr}}}
\newcommand{\AHdr}{$\aHdr$\xspace}
\newcommand{\aHdrF}{{\aHdr^{f}}}
\newcommand{\aHdrB}{{\aHdr^{b}}}
\newcommand{\AHdrF}{$\aHdrF$\xspace}
\newcommand{\AHdrB}{$\aHdrB$\xspace}
\newcommand{\onion}{O}
\newcommand{\Onion}{$\onion$\xspace}
\newcommand{\createAHdr}{\mbox{{\sc create\_ahdr}}}
\newcommand{\CreateAHdr}{$\createAHdr$\xspace}
\newcommand{\getFS}{\mbox{{\sc proc\_ahdr}}}
\newcommand{\GetFS}{$\getFS$\xspace}

\newcommand{\addLayer}{\mbox{{\sc add\_layer}}}
\newcommand{\AddLayer}{$\addLayer$\xspace}
\newcommand{\removeLayer}{\mbox{{\sc remove\_layer}}}
\newcommand{\RemoveLayer}{$\removeLayer$\xspace}

\providecommand{\aheader}{\mbox{a-header}\xspace}   \providecommand{\Aheader}{\mbox{A-header}\xspace}   

We now describe the details of our protocol, focusing on sender anonymity. We
begin with notation (Section~\ref{sec:notation}) and initialization
requirements (Section~\ref{sec:initialization}). We then describe the
establishment of anonymous communication sessions
(Section~\ref{sec:setupphase}) and data transmission
(Section~\ref{sec:datatransmission}).

\subsection{Notation}
\label{sec:notation}
Let $k$ be the security parameter used in the protocol.
For evaluation purposes we consider $k = 128$.
$\mathcal{G}$ is a prime order cyclic group of order  $q$ ($q\sim2^{2k}$), which satisfies the Decisional Diffie-Hellman Assumption. $\mathcal{G}^*$ is the set of non-identity elements in $\mathcal{G}$ and $g$ is a generator of $\mathcal{G}$. Throughout this section we use the multiplicative notation for $\mathcal{G}$.

Let $r$ be the maximum length of a path, \ie the maximum number of nodes on a path, including the destination. 
We denote the length of an FS as $|FS|$ and the size of an \aheader block, containing an FS and a MAC of size $k$, as $c=|FS| + k$. 

\name uses the following cryptographic primitives:
\begin{itemize}[leftmargin=*]
\item $\MAC: \{0,1\}^k \times \{0,1\}^*\rightarrow\{0,1\}^k$:  Message
Authentication Code (MAC) function. 
\item $\PRG, \PRGTWO, \PRGTHREE: \{0,1\}^k \rightarrow \{0,1\}^{rc}$: Three cryptographic pseudo-random generators.
\item $\PRP: \{0,1\}^k \times \{0,1\}^a \rightarrow \{0,1\}^a$: A pseudo-random permutation, implementable as a block cipher. The value of $a$ will be clear from the context.

\item $\ENC: \{0,1\}^k \times \{0,1\}^k \times \{0,1\}^{mk} \rightarrow \{0,1\}^{mk}$: Encryption function, with the second parameter being the Initialization Vector (IV) (\eg stream cipher in CBC mode). $m$ is a positive integer denoting the number of encrypted blocks.
\item $\DEC: \{0,1\}^k \times \{0,1\}^k \times \{0,1\}^{mk} \rightarrow \{0,1\}^{mk}$: Decryption 
function, inverse of $\ENC$. 
\item $h_{\mathsf{op}}: \mathcal{G}^* \rightarrow \{0,1\}^k$: a family of hash functions used to key $\mathsf{op}$, with $\mathsf{op} \in \{\MAC, \PRG, \PRGTWO, \PRP, \ENC, \DEC\}$.
\end{itemize}
We denote by \randomgen$(a)$ a function that generates a new uniformly random string of length $a$.

Furthermore, we define the notation for bit strings.
$0^a$ stands for a string of zeros of length $a$.
$|\sigma|$ is the length of the bit string $\sigma$. $\sigma_{[a\ldots{}b]}$ represents a substring of $\sigma$ from bit $a$ to bit 
$b$, with sub-index $a$ starting from 0; $\sigma_{[a\ldots{}end]}$ indicates the substring of $\sigma$ from bit $a$ till the end.  $\varepsilon$ is the empty string. $\sigma\concat \sigma'$ is the concatenation
of string $\sigma$ and string $\sigma'$. We summarize protocol notation and typical values for specific parameters in Table~\ref{tab:notation}. 

In the following protocol description, we consider a source $S$ communicating with a destination $D$ using forward path $p^f$ traversing nodes $n^f_0, n^f_1, \ldots, n^f_{l^f-1}$ and backward path $p^b$ traversing nodes $n^b_0, n^b_1, \ldots, n^b_{l^b-1}$, with $l^f, l^b \le r$, where $n^f_0$ and $n^b_{l^b-1}$ are the nodes closest to the source. Without loss of generality, we let the last node on the forward path $n_{l^f-1}^f = D$ and refer to the destination by these two notations interchangeably.
In general we use $dir \in \{f,b\}$ as superscripts to distinguish between notation referring to the forward and backward path, respectively.
Finally, to avoid redundancy, we use $\{sym_i^{dir}\}$ to denote $\{sym_i^{dir} | 0\le i \le l^{dir} - 1\}$, where $sym$ can be any symbol.

\begin{table}
	\centering
	\renewcommand{\arraystretch}{1.5}
	\scriptsize
	\begin{tabular}{c|m{6.5cm}}
		\hline
		Term & Definition \\\hline
		$k$ & Security parameter (length of keys and MACs). $k=128$ bits (16 B).\\
		$|FS|$ & Length of a forwarding segment (FS). $|FS|=256$ bits (32 B).\\
		$c$ & Length of a typical block made of an FS and a MAC. $c=|FS| + k =384$ bits (48 B) .\\
		$r$ & Maximum path length, including the destination. From our evaluation, $r=7$.\\
		$S, D$ & Source and destination. \\
		$p^f, p^b$ & The forward path (from S to D) and the backward path (from D to S).\\
		$l^f, l^b$ & Lengths of the forward and backward path ($l$, when it is clear from the context to which path it refers). From our evaluation, $1 \le  l \le 7$.\\
		$n_i^f, n_j^b$ & The $i$-th node on the forward path and the $j$-th node on the backward path, with $0 \le i < l^f$ and $0 \le j < l^b$. \\
		$g^{x_n}, x_n$ & Public/private key pair of node $n$. \\
		$s_i^f$ & Secret key established between S and node $n_i^f$. \\
		$R$ & Routing information, which allows a node to forward a packet to the next hop. \\
		$\cmnHdr$ & Common header. First three fields of both setup packets and data packets (see Figure~\ref{fig:setup_pkts}).        		\\
		$\sphinxHdr, \sphinxPayload$ & Sphinx header and payload. \\
		$\fsPayload$ & FS payload, used to collect the FSes during the setup phase. \\
		$\aHdr$ & Anonymous header, used for every data packet. It allows each node on the path to retrieve its FS. \\
		$\onion$ & Onion payload, containing the data payload of data packets. \\
		$\expTime$ & Expiration time, included in each FS. \\
		\hline
	\end{tabular}
	\normalsize
	\caption{Protocol notation and typical values (where applicable).}\label{tab:notation}
\end{table}

\subsection{Initialization}
\label{sec:initialization}

Suppose that a source $S$ wishes to establish an anonymous session with a public destination $D$.
First, $S$ anonymously obtains (from the underlying network) paths in both directions: 
a forward path $p^f = \{R^f_0, R^f_1, \cdots, R^f_{l^f - 1}\}$ from $S$ to $D$
and a backward path $p^b = \{R^b_0, R^b_1, \cdots, R^f_{l^b - 1}\}$ from $D$ to $S$.
$R^{dir}_{i}$ denotes the 
routing information needed by the node $n_i^{dir}$ to forward a packet. $S$ also anonymously 
retrieves and verifies a set of public keys $g^{x_{n_i^{dir}}}$ for the node $n^{dir}_i$ on path $p^{dir}$ (see Section~\ref{sec:network_model}). Note that $g^{x_D}$ is also included in the above set (as $n_{l^f - 1}^{f} = D$).
Finally, $S$ generates a random
DH public/private key pair for the session: $x_{S}$ and $g^{x_{S}}$. The per-session public key $g^{x_{S}}$
is used by the source
to create shared symmetric keys with nodes on the paths later in the setup phase.
$S$ locally stores $\left\{  \left(x_S, g^{x_{S}} \right), \left\{ g^{x_{n_i^{dir}}}\right\} , p^{dir} \right\}$, and uses these values for the setup phase.

\subsection{Setup Phase}
\label{sec:setupphase}
As discussed in Section~\ref{sec:overview}, 
in the setup phase, \name uses two \sphinx packets, which we denote by \setupfirst and \setupsecond, to traverse all nodes on both forward and backward paths and establish per-session state with every intermediate node, without revealing $S$'s network location.
For $S$ to collect the generated per-session state from each node, both \sphinx packets contain an empty FS payload into which each intermediate node can insert its FS, but is not able to learn anything about, or modify, previously inserted FSes.

\subsubsection{\sphinx Overview}
\label{sec:sphinx}
\sphinx~\cite{Danezis2009} is a provably-secure mix protocol.
Each \sphinx packet allows a source node to establish a set of symmetric keys, one for each node on the path through which packets are routed. These keys enable each node to check the
header's integrity, onion-decrypt the data payload, and retrieve the information to route the packet.
Processing \sphinx packets involves expensive asymmetric cryptographic operations, thus
\sphinx alone is not suitable to support high-speed anonymous communication.

\paragraph{\sphinx packets}
A \sphinx packet is composed of a \sphinx header \SphinxHdr and a \sphinx payload \SphinxPayload. 
The \SphinxHdr contains a group element $\sphinxPerNodePrivKey^{dir}$ that is re-randomized at each hop. Each $\sphinxPerNodePrivKey^{dir}$ is used as $S$'s ephemeral public key in a DH key exchange with node $n_i^{dir}$. From this DH exchange, node $n_i^{dir}$ derives a shared symmetric key $s_i^{dir}$, which it uses to process the rest of the \SphinxHdr and mutate $\sphinxPerNodePrivKey^{dir}$.
The rest of the \SphinxHdr is an onion-encrypted data structure, with each layer containing routing information and a MAC. The routing information indicates to which node the packet should be forwarded to next, and the MAC allows to check the header's integrity at the current node.
The \sphinx payload \SphinxPayload allows end hosts to send confidential content to each other. Each intermediate node processes \SphinxPayload by using a pseudo-random permutation.
 
\paragraph{\sphinx core functions}
We abstract the \sphinx protocol into the following six functions: 
\begin{itemize}[leftmargin=*]
	\item \sphinxGenSphinxHdr. The source uses this function to generate two \sphinx headers, \SphinxHdrSend and \SphinxHdrRecv, for the forward and backward path, respectively.
	It also outputs
	the symmetric keys $\{s_i^{dir}\}$, each established with the corresponding node's public key $g^{x_{n^{dir}_i}}$.
	\item \sphinxGenSphinxPayloadSend. The function allows the source to generate an onion-encrypted 
	payload \SphinxPayloadSend encapsulating confidential data to send to the destination. 
    \item \SphinxUnwrapSphinxPayloadSend. The function removes the last encryption layer added by \SphinxGenSphinxPayloadSend, and allows the destination to decrypt the \SphinxPayloadSend.
    \item \sphinxGenSphinxPayloadRecv. The function enables the destination to cryptographically wrap a data payload into \SphinxPayloadRecv before sending it to the source. 
   	\item \SphinxUnwrapSphinxPayloadRecv. 
		The function allows the source to recover the plaintext of the payload that the destination sent.
	\item \sphinxProcSphinxPkt. Intermediate nodes use this function to process a \sphinx packet,
	and establish symmetric keys shared with the source. 
	The function takes as inputs the packet $(\sphinxHdr, \sphinxPayload)$,
	 and the node's DH public key $g^{x_{n_i^{dir}}}$.
	 	 The function outputs the processed \sphinx packet $(\sphinxHdr', \sphinxPayload')$ and the established symmetric key $s_{i}^{dir}$.
		
\end{itemize}
\subsubsection{Forwarding Segment}
\label{sec:fs}
We extend Sphinx to allow each node to create a Forwarding Segment (FS) and add it to a data structure we name FS payload (see below).
An FS contains a node's per-session state, which consists of a secret key $s$ shared with the source, a routing segment $R$, and the session's expiration time \ExpTime. To protect these contents, the FS is encrypted with a PRP keyed by a secret value $SV$ known only by the node that creates the FS. A node seals and unseals its state using two opposite functions: \FsCreate and \FsOpen. They are defined as follows:
\begin{equation}
\label{eq:fs_definition}
\begin{aligned}
\FS & {}= \fsCreate(SV, s, R, \expTime) = \\
& = \PRP(h_\PRP(SV); \{s \concat R \concat \expTime\})
\end{aligned}
\end{equation}

\begin{equation}
\begin{aligned}
\{s \concat R \concat \expTime\} & {}= \fsOpen(SV, \FS) \\
& = \PRP^{-1}(h_{\PRP}(SV); \FS)
\end{aligned}
\end{equation}

\subsubsection{FS Payload} \label{sec:fs_payload}
At the end of each \name setup packet is a data structure we call FS payload (see Figure~\ref{fig:setup_pkts}). The FS payload is an onion-encrypted construction that allows intermediate nodes to add their FSes as onion-layers. 

Processing the FS payload leaks no information about the path's length or about an intermediate node's position on the path. All FS payloads are padded to a fixed length, which is kept constant by dropping the right number of trailing bits of the FS payload before an FS is added to the front. Moreover, new FSes are always added to the beginning of the FS payload, eliminating the need for intermediate nodes to know their positions in order to process FS payloads.

An FS payload also provides both secrecy and integrity for the FSes it contains. Each node re-encrypts the FS payload after inserting a new FS and computes a MAC over the resulting structure. Only the source, with symmetric keys
shared with each node on a path, can retrieve all the FSes from the FS payload and verify their integrity.

\paragraph{Functions}
There are three core functions for the FS payload: \FsPayloadInit, \AddFs, and \RetrieveFs.

\emph{\FsPayloadInit}.
A node initializes an FS payload by using a pseudo-random generator keyed with a symmetric key $s$ to generate $rc$ random bits:
\begin{eqnarray}
\fsPayload = \PRGTWO(h_{\PRGTWO}(s))
\end{eqnarray}
where $c=|FS| + k$ is the size of a basic block of the FS payload (consisting of an FS and a MAC).

\emph{\AddFs}. 
Each intermediate node uses \AddFs to insert its FS
into the payload, as shown in Algorithm~\ref{alg:add_fs}. First, the trailing $c$ bits of the current FS payload, which are padding bits containing no information about previously added FSes, are dropped, and then the FS
is prepended to the shortened FS payload. The result is encrypted using a stream cipher (Line~\ref{alg:add_fs:encrypt}) and MACed (Line~\ref{alg:add_fs:mac}).
Note that no node-position information is required in \AddFs, and verifying that the length of the FS payload remains unchanged is straightforward.

\begin{algorithm}
	\begin{algorithmic}[1]
		\Procedure{add\_fs}{}
		\Statex Input: $s$, $\FS$, $P_{in}$
		\Statex Output: $P_{out}$
		\State $P_{tmp} \gets \left\{\FS \concat {P_{in}}_{[0 .. (r-1)c-1]}\right\}$
		\label{alg:add_fs:encrypt}
		\StatexIndent[3] $\;\,\vphantom{}\oplus \PRG(h_{\PRG}(s))_{[k .. end]}$
		\State $\alpha \gets \MAC(h_{\MAC}(s); P_{tmp})$
		\State $P_{out} \gets \alpha \concat P_{tmp}$
		\label{alg:add_fs:mac}
		\EndProcedure
	\end{algorithmic}
	\caption{Add FS into FS payload.}
	\label{alg:add_fs}
\end{algorithm}

\emph{\RetrieveFs}.
The source uses this function to recover all FSes $\{FS_i\}$
inserted into an FS payload \FsPayload. 
\RetrieveFs starts by recomputing the discarded trailing bits (Line~\ref{alg:retrieve_fses:pad}) and obtaining a complete payload $P_{full}$. Thus, intuitively, this full payload is what would remain if no nodes dropped any bits before inserting a new FS. 
Afterwards, the source retrieves the FSes
from $P_{full}$ in the reverse order in which they were added by \AddFs (see lines~\ref{alg:retrieve_fses:mac} and \ref{alg:retrieve_fses:decrypt}). 

\begin{algorithm}
	\begin{algorithmic}[1]
		\Procedure{retrieve\_fses}{}
		\Statex Input: $\fsPayload$, $s$, $\{s_i\}$
		\Statex Output: $\{FS_i\}$		\State $P_{init}  \gets  \fsPayloadInit(s)$
		\State $\psi \gets {P_{init}}_{[(r-l)c .. rc - 1]}$
		\StatexIndent[2] $\vphantom{0} \oplus \PRG(h_{\PRG}(s_0))_{[(r-l+1)c .. end]} \concat 0^{c}$
		\StatexIndent[2] $\vphantom{0} \oplus \PRG(h_{\PRG}(s_1))_{[(r-l+2)c .. end]} \concat 0^{2c}$
		\StatexIndent[3] $\cdot\cdot\cdot $
		\StatexIndent[2] $\vphantom{0} \oplus \PRG(h_{\PRG}(s_{l-2}))_{[(r-1)c .. end]} \concat 0^{(l-1)c}$ \label{alg:retrieve_fses:pad}
		\State $P_{full} = \fsPayload \concat \psi$
		\For{$i \gets (l-1), \dots, 0$}
		\State \textbf{check} ${P_{full}}_{[0 .. k-1]} = \vphantom{0}$ \label{alg:retrieve_fses:mac}
		\StatexIndent[5] $\MAC(h_{\MAC}(s_i); {P_{full}}_{[k .. rc -1]})$
		\State $P_{full} \gets P_{full} \oplus (\PRG(h_{\PRG}(s_i)) \concat 0^{(i+1)c})$
		\State $\FS_i \gets {P_{full}}_{[k .. c - 1]}$ \label{alg:retrieve_fses:decrypt}
\vspace{5pt}
		\State $P_{full} \gets {P_{full}}_{[c .. end]}$
		\EndFor
		\EndProcedure
	\end{algorithmic}
	\caption{Retrieve FSes from FS payload.}
	\label{alg:retrieve_fses}
\end{algorithm}

\subsubsection{Setup Phase Protocol Description}
\paragraph{Source processing}
With the input
$$
I = \left\{  \left(x_S, g^{x_{S}} \right), \left\{ g^{x_{n_i^{dir}}}\right\} , p^{dir} \right\}
$$
the source node $S$ bootstraps a session setup in 5 steps:
\begin{enumerate}[leftmargin=*]
    \item $S$ selects the intended expiration time \ExpTime for the session and
        specifies it in the common header \CmnHdr (see
        Section~\ref{sec:packetstructure}).\footnote{\ExpTime must not become
        an identifier that allows matching packets of the same flow across
        multiple links. Since \ExpTime does not change during setup packet forwarding, a coarser granularity (e.g., 10s) is desirable. 
                In addition, the duration of the session should also have only a
restricted set of possible values (\eg 10s, 30s, 1min, 10min) to avoid matching
packets within long sessions. For long-lived connections, the
source can create a new session in the background before expiration of the
previous one to avoid additional latency.}
	\item $S$ generates the send and the reply \sphinx headers by:
	\begin{eqnarray}
	\{\sphinxHdrSend, \sphinxHdrRecv\} = \sphinxGenSphinxHdr(I, \cmnHdr)
	\end{eqnarray}
	The common header \CmnHdr (see Figure~\ref{fig:setup_pkts}) is passed to the function to extend the per-hop integrity protection of Sphinx over it. \SphinxGenSphinxHdr also produces the symmetric keys shared with each node on both paths $\{s_i^{dir}\}$.
	
	\item In order to enable the destination $D$ to reply, $S$ places the reply \sphinx header \SphinxHdrRecv into the \sphinx payload:
	\begin{eqnarray}
	\sphinxPayload^f = \sphinxGenSphinxPayloadSend(\{s_i^f\}, \sphinxHdrRecv)
	\end{eqnarray}
	
	\item $S$ creates an initial FS payload $\fsPayload^f = \fsPayloadInit(x_S)$.	
	\item $S$ composes $\mbox{\setupfirst}=\{\cmnHdr\concat \sphinxHdrSend\concat\sphinxPayload^f\concat\fsPayload^f\}$ and sends it to the first node on the forward path $n_0^f$.
\end{enumerate}
	
\paragraph{Intermediate node processing}
An intermediate node $n_i^{f}$ receiving a packet $\mbox{\setupfirst}=\{\cmnHdr\concat \sphinxHdrSend\concat \sphinxPayload^f \concat \fsPayload^f\}$ processes it as follows:
\begin{enumerate}[leftmargin=*]
 \begin{sloppypar}
    \item $n_i^{f}$ first processes $\sphinxHdrSend$ and $\sphinxPayload^f$ in \setupfirst according to the \sphinx protocol (using \sphinxProcSphinxPkt). As a result $n_i^{f}$ obtains the established symmetric key $s_i^f$ shared with $S$, the processed header and payload $(\sphinxHdrSend', {\sphinxPayload^f}')$ as well as the routing information $R_i^f$. During this processing the integrity of the \CmnHdr is verified.
 \end{sloppypar}

	\item $n_i^f$ obtains $\expTime$ from \CmnHdr and checks that $\expTime$ is not expired.
	$n_i^f$ also verifies that $R_i^f$ is valid.

	\item $n_i^f$ generates its forwarding segment $\FS_{i}^f$ by using its local symmetric key $SV_i^f$ to 
	encrypt $s_i^f$, $R_i^f$, and $\expTime$ (see Equation~\ref{eq:fs_definition}):
	\begin{equation}
	\FS_i^f = \fsCreate(SV_i^f, s_i^f, R_i^f, \expTime)
	\end{equation}

	\item $n_i^f$ adds its $\FS_i^f$
	into the FS payload $\fsPayload^f$.
	\begin{equation}
	{\fsPayload^f}' = \addFs(s^f_i, FS_i^f, \fsPayload^f)
	\end{equation}
	
	\item Finally node $n_i^f$ assembles the processed packet $\mbox{\setupfirst} = \{\cmnHdr\concat \sphinxHdrSend'\concat{\sphinxPayload^f}'\concat {\fsPayload^f}'\}$ and routes it to the next node according to the routing 
	information $R^f_i$.
	
\end{enumerate}

\paragraph{Destination processing}
\label{sec:dst_proc}
As the last node on the forward path, $D$ processes \setupfirst in the same way as the previous nodes. It first processes the \sphinx packet in \setupfirst and derives a symmetric key $s_{D}$ shared with $S$,
and then it encrypts per-session state, including $s_D$, into $FS_D$, and inserts $FS_D$ into the FS payload.

After these operations, however, $D$ moves on to create the second setup packet \setupsecond as follows:
\begin{enumerate}[leftmargin=*]
	\item $D$ retrieves the \sphinx reply header using the symmetric key $s_D$:
	\begin{equation}
	\sphinxHdrRecv = \sphinxUnwrapSphinxPayloadSend(s_D, \sphinxPayload^f)
	\end{equation}
	\item $D$ places the FS payload $\FsPayload^f$ of \setupfirst into the \sphinx payload \SphinxPayloadRecv of \setupsecond (this will allow $S$ to get the FSes $\{\FS_i^f\}$):
	\begin{equation}
	\sphinxPayload^b = \sphinxGenSphinxPayloadRecv(s_D, \fsPayload^f)
	\end{equation} 
	Note that since $D$ has no knowledge about the keys $\{s_i^f\}$ except for $s_D$, $D$ learns nothing about the other FSes in the FS payload.
	\item $D$ creates a new FS payload $\fsPayloadRecv = \fsPayloadInit(s_D)$ to collect the FSes along the backward path.
	\item $D$ composes $\mbox{\setupsecond} = \{\cmnHdr \concat \sphinxHdrRecv \concat \sphinxPayload^b \concat \fsPayloadRecv\}$ and sends it to the first node on the backward path, $n_0^b$. 
\end{enumerate}

The nodes on the backward path process \setupsecond in the exact same way nodes on the forward path processed \setupfirst. Finally \setupsecond reaches the source $S$ with FSes $\{\FS_i^b\}$ added to the FS payload.

\paragraph{Post-setup processing}
Once $S$ receives \setupsecond it extracts all FSes, \ie $\{\FS_i^f\}$ and $\{\FS^b_i\}$, as follows:
\begin{enumerate}[leftmargin=*]
	\item $S$ recovers the FS payload for the forward path $\fsPayload^{f}$ from \SphinxPayloadRecv:
	\begin{equation}
	\fsPayload^f = \sphinxUnwrapSphinxPayloadRecv(\{s_i^b\}, \sphinxPayload^b)
	\end{equation}
	\item $S$ retrieves the FSes for the nodes on the forward path $\{\FS_i^f\}$:
	\begin{equation}
	\{\FS_i^f\} = \retrieveFs(\{s_i^f\}, \fsPayload^f)
	\end{equation}
	\item $S$ directly extracts from \FsPayloadRecv the FSes for the nodes on the backward path $\{\FS_i^b\}$:
	\begin{equation}
	\{\FS_i^b\} = \retrieveFs(\{s_i^b\}, \fsPayloadRecv)
	\end{equation}
\end{enumerate}

With the FSes for all nodes on both paths, $\big\{\FS_i^f\big\}$ and $\big\{\FS_i^b\big\}$, $S$ is ready to start the data transmission phase.

\subsection{Data Transmission Phase}
\label{sec:datatransmission}
\begin{figure}[!htbp]
	\centering
	\includegraphics[width=8cm]{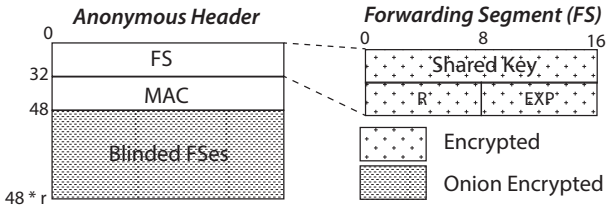}
	\caption{Format of a \name anonymous header with details of a forwarding segment (FS).}
	\label{fig:a_header}
\end{figure}

Each \name data packet contains an anonymous header \AHdr and an onion-encrypted payload \Onion as shown in Figure~\ref{fig:setup_pkts}. Figure~\ref{fig:a_header} illustrates the details of an \AHdr. The \AHdr allows each intermediate node along the path to retrieve its per-session state in the form of an FS and process the onion-encrypted data payload. All  processing of data packets in \name only involves symmetric-key cryptography, therefore supporting fast packet processing.

At the beginning of the data transmission phase, $S$ creates two \aheader{}s, one for the forward path (\AHdrF) and one for the backward path (\AHdrB), by using FSes collected during the setup phase. \AHdrF enables $S$ to send data payloads to $D$. To enable $D$ to transmit data payloads back, $S$ sends \AHdrB as payload in the first data packet. If this packet is lost, the source would notice from the fact that no reply is seen from the destination. If this happens the source simply resends the backward \aheader using a new data packet.

\subsubsection{Anonymous Header}
\label{sec:aheader}
Like an FS payload, an \aheader is an onion-encrypted data structure that contains FSes. It also offers the same guarantees, \ie secrecy and integrity, for the individual FSes it contains, for their number and for their order. Its functionalities, on the other hand, are the inverse: while the FS payload allows the source to collect the FSes added by intermediate nodes, the \aheader enables the source to re-distribute the FSes back to the nodes for each transmitted data packet.

\vspace{1.5em}
\paragraph{Functions} 
The life cycle of \aheader{}s consists of two functions: the header construction (\CreateAHdr) and the header processing (\GetFS). We begin with the description of \GetFS since it is simpler, and its helps understand the construction of \CreateAHdr.\GetFS allows each intermediate node to verify the integrity of an
incoming \aheader, and to check that the corresponding session has not expired.
\GetFS also retrieves the key $s$ shared with the source, as well as the
routing information $R$, from the FS of the node invoking the function.
Finally, \GetFS also returns the processed  header $\aHdr'$, which will be used
by the next hop. The details of this function can be seen in
Algorithm~\ref{alg:get_fs}.

\begin{algorithm}
	\begin{algorithmic}[1]
		\Procedure{proc\_ahdr}{}
		\Statex Input: $SV$, \AHdr
		\Statex Output: $s$, $R$, $\aHdr'$
		\State $\{\FS \concat \gamma \concat \beta\} \gets \aHdr$ \label{alg:get_fs:fs}
		\State $\{s \concat R \concat \expTime\} \gets \fsOpen(SV, \FS)$  \label{alg:get_fs:key}
		\State \textbf{check} $\gamma = \MAC(h_{\MAC}(s); \FS \concat \beta)$
		\State \textbf{check} $t_{curr} < \expTime\quad$
		\State $\aHdr' \gets \{\beta \concat 0^c \} \oplus \PRGTHREE(h_{\PRGTHREE}(s))$ \label{alg:get_fs:decrypt}
		\EndProcedure
	\end{algorithmic}
	\caption{Process an \aheader.}
	\label{alg:get_fs}
\end{algorithm}

\vspace{1.5em}
\begin{sloppypar}
Our \aheader construction resembles the \sphinx packet header construction~\cite{Danezis2009}.
For each path (forward and backward), \CreateAHdr enables $S$ to create an \aheader given the keys $\{s_i\}$ shared with each node on that path, and given the forwarding segments $\{FS_i\}$ of those nodes. All these keys and FSes are obtained during the setup phase (see Section~\ref{sec:setupphase}). The details are shown in Algorithm~\ref{alg:construct_aheader}.
In essence, \CreateAHdr is equivalent to a series of \GetFS iterations performed in reverse.
Initially, the paddings $\phi$ are computed, each of which is the leftmost part of an \aheader that results from the successive encryptions of the zero-paddings added in \GetFS ($\phi_0$ is the empty string since no padding has been added yet).
Once the last padding is computed (the one for the \aheader received by the last hop, $\phi_{l-1}$), the operations in \GetFS are reversed, obtaining at each step the \aheader{}s as will be received by the nodes, from the last to the first. This also allows the computation of the per-hop MACs.
\end{sloppypar}

\begin{algorithm}
	\begin{algorithmic}[1]
		\Procedure{create\_ahdr}{}
		\Statex Input: $\{s_i\}$, $\{FS_i\}$
		\Statex Output: $(\FS_0, \gamma_0, \beta_0)$
		\State $\phi_0 \gets \varepsilon$ 
		\For{$i \gets 0,\dots, l-2$}
		\State $\phi_{i+1} \gets (\phi_{i} \concat 0^{c})$
		\StatexIndent[4] $\vphantom{0}\oplus \left\{ \PRGTHREE(h_{\PRGTHREE}(s_{i}))_{[(r-1-i)c .. end]} \right\}$
		\EndFor		
		\State $\beta_{l-1} \gets \mbox{\randomgen}\left((r-l)c\right) \concat \phi_{l-1} $ \label{alg:create_aheader_start}
		\State $\gamma_{l-1} \gets \MAC(h_{\MAC}(s_{l-1}); \FS_{l-1} \concat \beta_{l-1})$
		\For{$i \gets (l-2), \ldots, 0$}
		\State $\beta_{i} \gets \left\{\FS_{i+1} \concat \gamma_{i+1} \concat {\beta_{i+1}}_{[0 .. (r-2)c-1]}\right\} $
		\StatexIndent[4] $\vphantom{0} \oplus \PRGTHREE(h_{\PRGTHREE}(s_i))_{[0 .. (r-1)c-1]} $ \label{alg:create_aheader:beta}
		\State $\gamma_{i} \gets \MAC(h_{\MAC}(s_{i}); \FS_i \concat \beta_{i})$ \label{alg:create_aheader:gamma}
		\EndFor
		\EndProcedure
	\end{algorithmic}
	\caption{Anonymous header construction.}
	\label{alg:construct_aheader}
\end{algorithm}

\subsubsection{Onion Payload}
\label{sec:onion_payload}
\name data payloads are protected by onion encryption. To send a data payload to the destination, the source adds a sequence of encryption layers on top of the data payload, one for each node on the forward path (including the destination). As the packet is forwarded, each node removes one layer of encryption, until the destination removes the last layer and obtains the original plaintext. 

To send a data payload back to the source, the destination adds only one layer of encryption with its symmetric key shared with the source. As the packet is forwarded, each node on the backward path re-encrypts the payload until it reaches the source. With all the symmetric keys shared with nodes on the backward path, 
the source is capable of removing all encryption layers, thus obtaining the original data payload sent by the destination.

\paragraph{Functions}
Processing onion payloads requires the following two functions: \AddLayer and \RemoveLayer.

\emph{\AddLayer}.
The function's full form is:
\begin{equation}
\{O', IV'\} = \addLayer(s, IV, O)
\end{equation}
Given a symmetric key $s$, an initial vector $IV$, and an input onion payload $O$, \AddLayer performs two tasks. First, \AddLayer encrypts $O$ with $s$ and $IV$:
\begin{equation}
O' = \ENC(h_{\ENC}(s); IV; O)
\end{equation} 
Then, to avoid making the IV an identifier across different links, \AddLayer mutates the $IV$ for the next node:
\begin{equation}
IV' = \PRP(h_{\PRP}(s); IV)
\end{equation}

\emph{\RemoveLayer}. The function is the inverse of \AddLayer, decrypting the onion payload at each step, and mutating the $IV$ using the inverse permutation $\PRP^{-1}$ keyed with $h_{\PRP}(s)$. Its full form is the following:
\begin{equation}
\{O', IV'\} = \removeLayer(s, IV, O)
\end{equation}

\subsubsection{Initializing Data Transmission}
To start the data transmission session, $S$ generates \AHdrF and \AHdrB as follows:\begin{eqnarray}
\aHdrF &=& \createAHdr(\{s_i^f\}, \{\FS_i^f\})\\
\aHdrB &=& \createAHdr(\{s_i^b\}, \{\FS_i^b\})
\end{eqnarray}
$S$ then sends $\aHdrB$ to $D$ as payload of the first data packet (which uses $\aHdrF$), as specified in the following section.

\subsubsection{Data Transmission Protocol Description}
\label{sec:DataTransmissionProtocolDescription}
\paragraph{Source processing}
With \AHdrF, $S$ can send a data payload $M$ with the following steps:
\begin{enumerate}[leftmargin=*]
	\item $S$ ensures that the session is not expired by checking that the current time $t_{curr} < \expTime$.
	\item $S$ creates an initial $IV$. With the shared keys $\{s_i^f\}$, $S$ onion encrypts the data payload $M$ by setting $O_{l^f} = M$ and $IV_{l^f}=IV$ and computing the following for $i \gets (l^f - 1) .. 0$:
	\begin{equation}
	\{O_{i}, IV_{i}\} = \addLayer(s_{i}, IV_{i+1}, O_{i+1})
	\end{equation}
	
	\item $S$ places $IV_{0}$ in the common header \CmnHdr.
	\item $S$ sends out the resulting data packet $\{ \cmnHdr, \aHdrF, O_0\}$. 
\end{enumerate}

\paragraph{Processing by intermediate nodes}
Each intermediate node $n_i^f$ on the forward path processes a received data packet of the form $\{\cmnHdr, \aHdrF, O\}$ with its local secret key $SV_i^f$ as follows:
\begin{enumerate}[leftmargin=*]
	\item $n_i^f$ retrieves the key $s^f_i$ shared with $S$ and the routing information $R^f_i$ from $\aHdrF$:
	\begin{equation}
	\{s_i^f, R_i^f, \aHdrF'\} = \getFS(SV_i^f, \aHdrF)
	\end{equation}
	\GetFS also verifies the integrity of $\aHdr$, and checks that the session has not expired.

	\item $n_i^f$ obtains $IV$ from $\cmnHdr$ and removes one layer of encryption from the data payload:
	\begin{equation}
	\{O', IV'\} = \removeLayer(s_i^f, IV, O)
	\end{equation}
	\item $n_i^f$ updates the IV field in \CmnHdr with $IV'$.
	\item $n_i^f$ sends the resulting packet $\{\cmnHdr', \aHdrF', O'\}$ to the next node according to $R_i^f$.	
\end{enumerate}

The above procedures show that the intermediate node processing requires only symmetric-cryptography operations.

\paragraph{Destination processing}
$D$ processes incoming data packets as the intermediate nodes. Removing the last encryption layer from the onion payload $D$ obtains the original data payload $M$ sent by $S$.
Additionally, for the first data packet $D$ retrieves \AHdrB from the payload, and stores the $\{s_D, R_0^b, \aHdrB\}$ locally so that $D$ can retrieve \AHdrB when it wishes to send packets back to $S$.

\paragraph{Processing for the backward path} 
Sending and processing a \name packet along the backward path is the same as that for the forward path, with the exception of processing involving the data payload.
Because $D$ does not possess the symmetric keys that each node on the backward path shares with $S$, $D$ cannot onion-encrypt its payload.
Therefore, instead of \RemoveLayer, $D$ and the intermediate nodes use \AddLayer to process the data payload, and the source node recovers the data with \RemoveLayer.

\subsection{Nested Anonymous Header Construction}

As discussed in Section~\ref{sec:recv_anonymity}, the main difference of the protocols between sender anonymity and sender-receiver anonymity is that the latter requires nested \aheader{}s.
We present in detail the process of composing an \aheader with a nested \aheader in Algorithm~\ref{alg:create_nested_aheader}.

Constructing a new \aheader based on a nested \aheader $A$ has essentially the same procedures 
as constructing a normal \aheader from ASes, except for the initialization process and 
the size of the resulted \aheader. For the \aheader initialization in Line~\ref{alg:line:init} in Algorithm~\ref{alg:create_nested_aheader},
the nested \aheader $A$ is perpended to the random bits generated. Thus, when the last node $n_l^{dir}$ (RP)
decrypts the \aheader, $A$ is revealed to the node. For the size of the resulting \aheader, instead of $r$ for a
normal \aheader, the length of the generated \aheader with a nested \aheader is $2r$, doubling the bandwidth cost
incurred by the protocol headers.

\begin{algorithm}
	\begin{algorithmic}[1]
		\Procedure{create\_padding\_string\_nested}{}
		\Statex Input: $\{s_i\}$, $r$
		\Statex Output: $\phi_{l-1}$
		\State $\phi_0 \gets \varepsilon$ 
		\For{$0<i<l$}
		\State $\phi_i \gets (\phi_{i-1} \concat 0^{c}) \oplus$
		\State $\quad \quad$ $\left\{ \PRG(h_{\PRG}(s_{i-1}))_{[(2r-i)c .. 2rc]} \right\}$
		\EndFor
		\EndProcedure
		
		\Procedure{create\_anonymous\_header\_nested}{}
		\Statex Input: $\{s_i\}$, $\{FS_i\}$, $A$	
		\Statex Output: $\quad$ $(\FS_0, \gamma_0, \beta_0)$
		\State $\phi_{l-1} \gets$ {\sc create\_padding\_string\_nested}$(\{s_i\})$
		\State $\beta_{l-1} \gets \big\{ \left\{ A \concat \mbox{\randomgen}(c(r-l))\right\}$
		\StatexIndent[3] $\vphantom{0} \oplus \PRG(h_{\PRG}(s_{l-1}))_{[0 .. c(2r-l) - 1]} \big\} \concat \phi_{l-1}$
		\label{alg:line:init}
		\State $\gamma_{l-1} \gets \MAC(h_{\MAC}(s_{l-1}); \FS_{l-1} \concat \beta_{l-1})$
		\For{$i \gets (l-2), \ldots, 0$}
		\State $\beta_{i} \gets \left\{\FS_{i+1} \concat \gamma_{i+1} \concat {\beta_{i+1}}_{[0 .. c(2r-1)-1]}\right\} $
		\StatexIndent[4] $\vphantom{0} \oplus \PRG(h_{\PRG}(s_i))_{{[0 .. c(2r -l) - 1]}} $
		\State $\gamma_{i} \gets \MAC(h_{\MAC}(s_{i}); \FS_i \concat \beta_{i})$
		\EndFor
		\EndProcedure
	\end{algorithmic}
	\caption{Creating an \aheader with a nested \aheader{}.}
	\label{alg:create_nested_aheader}
\end{algorithm}

\section{Security Analysis}
\label{sec:security}

\newtheorem{lemma}{Lemma}

In this section, we first present formal proofs showing that \name satisfies the correctness, security, and integrity properties defined by Camenisch and Lysyanskaya~\cite{camenisch2005formal}. Then, we describe how \name defends against well-known de-anonymization attacks and meets the design goals of Section~\ref{sec:desired_properties}. We also present defenses against denial of service attacks. 

\subsection{Formal Proof of Security for HORNET Data Transmission Phase}
We prove \name{}'s data transmission phase realizes ideal onion routing functionalities in the Universal Composability (UC) framework~\cite{canetti2001universally}.
Conceptually, with an ideal onion routing protocol, adversaries have no access to the routing information or
the message within packets except for opaque identifiers that vary across links. 

As demonstrated by Camenisch and Lysyanskaya~\cite{camenisch2005formal}, to prove that a protocol conforms to
an ideal onion routing model, it is sufficient to show that 
the protocol provides four properties: \emph{correctness}, \emph{integrity}, \emph{wrap-resistance}, and 
\emph{security}.

\subsubsection{Correctness}
Proving the correctness property requires that \name protocol functions correctly in the absence of adversaries. 
A scrutiny of protocol description in Section~\ref{sec:protocol} should suffice.

\subsubsection{Integrity}
To prove the integrity property, we need to prove that an adversary cannot forge a message that can traverse 
more than $N$ uncompromised nodes, where $Q$ is a fixed upper bound for \name.
Equivalently, we demonstrate that 
an adversary, with significantly less than $2^k$ computation, can only produce a requisite message with a negligible
probability. In our proof, we choose $Q=r+1$.

Suppose that an adversary can construct a \name \aheader ($FS_0$, $\gamma{}_0$, $\beta{}_0$) that can succeed in 
traversing $r+1$ honest nodes $n_0$, $n_2$, $\dots$, $n_r$, without knowing secrets $SV_0$, $\dots$, $SV_r$. 
According to Algorithm~\ref{alg:construct_aheader}, 
$FS_r$, $\beta_r$, and $\gamma_r$ satisfy:
\begin{eqnarray}
	\gamma_r = MAC(h_{MAC}(PRP^{-1}(h_{PRP}(SV_r); FS_r)_{[0..c]}); \beta_r) \label{eq:int_cond_1}
\end{eqnarray}

For convenience, for $i\le j \le r-1$, we introduce the following notation:
\begin{eqnarray}
	\phi(SV, FS) &=&  PRP^{-1}\left(h_{PRP}\left(SV\right); FS\right)\\
	\rho(SV, FS) &=& PRG\left(h_{PRG}\left(\phi\left(SV,FS\right)\right)\right)\\
	\rho_i &=& \rho(SV_i, FS_i^*)\\
	\rho^{FS}_i &=& \left\{\rho_i\right\}_{[c(r-1-i) .. c(r-1-i)+l_{FS}-1]}\\
	\rho^{\gamma}_i &=& \left\{\rho_i\right\}_{[c(r-1-i) + l_{FS}  .. c(r-i)-1]}\\
	\rho^{\beta}_i &=& \left\{\rho_i\right\}_{[0 .. c(i+1)-1]} || 0^{c(r-1-i)}\\
	\rho^{c}_{i,j} &=& \left\{\rho_i\right\}_{[jc .. (j+1)c-1]} \label{eq:rho_c}
\end{eqnarray}
where $FS_i^*$ are defined recursively as follows:
\begin{eqnarray}
	FS_0^* &=& FS_0\\
	FS_i^* &=& FS_i \oplus \bigoplus_{j=0}^{i-1} \left\{\rho_i\right\}_{[c(j+i-1) .. c(j+i-1)+l_{FS}-1]}\label{eq:fs_start}
\end{eqnarray}
We observe that $FS_i^*$ is a function of $\{FS_j~|~\forall 0\le j \le i\}$ and $\{SV_j~|~\forall 0\le j \le i-1\}$. Accordingly,
$\rho^{FS}_i$, $\rho^{\gamma}_i$, and $\rho^{\beta}_i$ are all functions of $\{FS_j~|~\forall 0\le j \le i\}$ and $\{SV_j~|~\forall 0\le j \le i-1\}$.

With a detailed inspection of Algorithm~\ref{alg:construct_aheader}, we can express $FS_r$, $\beta_r$, and $\gamma_r$:
\begin{eqnarray}
	FS_r &=& \bigoplus_{i=0}^{r-1} \rho^{FS}_i \label{eq:FS_r}\\
	\gamma_r &=& \bigoplus_{i=0}^{r-1} \rho^{\gamma}_i\label{eq:gamma_r}\\
	\beta_r &=& \bigoplus_{i=0}^{r-1} \rho^{\beta}_i \label{eq:beta_r}\\
\end{eqnarray}  

With Equation~\ref{eq:FS_r}, \ref{eq:gamma_r}, \ref{eq:beta_r} and \ref{eq:int_cond_1}, we can prove the following 
lemma:
\begin{lemma}\label{lem:1}
	With less than $2^k$ work, an adversary can only distinguish $MAC(h_{MAC}(\phi(SV_r, FS_r)_{[0..c]}); \beta_r)$ from a random oracle with negligible probability.
\end{lemma}

\paragraph{Proof}
(Sketch) We will show that an adversary cannot find two sets of 
{\small
	$$(SV_0, \dots, SV_r, FS_0 \dots, FS_{r-1}) \neq (SV_0', \dots, SV_r', FS_0' \dots, FS_{r'-1})$$ 
}
that lead to the same value of $MAC(h_{MAC}(\phi(SV_r, FS_r)_{[0..c]}); \beta_r)$ with significant less than $2^k$ work. 
Assume that the adversary, with much less than $2^k$ work, finds two sets, 
$$(SV_0, \dots, SV_r, FS_0 \dots, FS_r) \neq (SV_0', \dots, SV_r', FS_0' \dots, FS_r')$$ 
that results in the same value of 
$$MAC(h_{MAC}(\phi(SV_r, FS_r)_{[0..c]}); \beta_r)$$
We will show the assumption leads to a contradiction.

Because $MAC$ is a random oracle, the only way for an attacker to distinguish the target function from a random oracle with much less than $2^k$ work is to ensure 
$$\phi(SV_r, FS_r)_{[0..c]}= \phi(SV_r', FS_r')_{[0..c]}$$
and $\beta_r = \beta_r'$. Because $PRP$ is a pseudo-random permutation and $h_{PRP}$ is collision resistant, 
we have $SV_r=SV_r'$.

Note that the last $c$ bits of $\beta_r$ and $\beta_r'$ are $\rho^{c}_{r-1, r-1}$ and ${\rho^{c}_{r-1, r-1}}'$ respectively.
Therefore, we have $\rho^{c}_{r-1, r-1} = {\rho_{r-1, r-1}^c}'$. According to Equation~\ref{eq:rho_c}, because $PRG$ 
is a pseudo-random generator, we have $SV_{r-1}=SV_{r-1}$ and $FS^*_{r-1} = {FS^*_{r-1}}'$. Hence, $\rho^c_{r-1, j} = {\rho^c_{r-1, j}}'$, $\forall 0\le j \le r-1$.

A careful calculation shows that the $c$ bits before the last $c$ bits in $\beta_r$ and $\beta_r'$ are
$\rho^c_{r-2, r-2} \oplus \rho^c_{r-1, r-2}$ and ${\rho^c_{r-2, r-2}}' \oplus {\rho^c_{r-1, r-2}}'$. Similarly,
we have $SV_{r-2} = {SV_{r-2}}'$ and $FS_{r-2}^* = {FS_{r-2}^*}'$.

Continuing the logic as above, we finally have $SV_{i}=SV_{i}'$ and $FS_{i}^* = {FS_{i}^*}'$, $\forall 0 \le i \le r-1$.
However, given Equation~\ref{eq:fs_start},  $SV_{i}=SV_{i}'$, and $FS_0^* = {FS_0^*}'$, we have $FS_i = FS_i'$, $\forall 0 \le i \le r-1$. This results in
{\small
	$$(SV_0, \dots, SV_r, FS_0 \dots, FS_{r-1}) = (SV_0', \dots, SV_r', FS_0' \dots, FS_{r'-1})$$
}
Therefore, we obtain a contradiction. \qed

We can substitute Equation~\ref{eq:FS_r}, \ref{eq:gamma_r}, and \ref{eq:beta_r} into Equation~\ref{eq:int_cond_1},
and rewrite the equation into:
\begin{eqnarray}
	\rho^{\gamma}_0= MAC(h_{MAC}(\phi(SV_r, FS_r)_{[0..c]}); \beta_r) \oplus \bigoplus_{i=1}^{r-1}\rho^{\gamma}_i \label{eq:wrap_contradict}
\end{eqnarray}
Because $MAC$ is not used in $\rho^{\gamma}_i$, the right side of Equation~\ref{eq:wrap_contradict} is a random oracle 
with respect to $SV_i$ and $FS_i$, $\forall 0\le i\le r-1$.

We can further simplify the notation by denoting $\rho^{\gamma}_0$ as $f_0(SV_0, FS_0)$ and the right side of Equation~\ref{eq:wrap_contradict} as 
$$f_1(FS_0, \dots, FS_{r-1}, SV_0, \dots, SV_{r-1})$$
Both $f_0$ and $f_1$ are random oracles with range $\{0,1\}^k$. As a result, by creating a \aheader traversing 
$r+1$ honest nodes, the adversary equivalently finds a solution to 
$$f_0(SV_0, FS_0)=f_1(FS_0, \dots, FS_{r-1}, SV_0, \dots, SV_{r-1})$$ which can only be solved
with negligible probability with significantly less than $2^k$ work.
Hence, with much less than $2^k$ work, the adversary can only generate a packet that traverse $r+1$ hops with 
negligible probability.

\subsubsection{Wrap-resistance}

To prove the wrap-resistance property, we show that given a data packet $(FS, \gamma, \beta, P)$, an adversary, with significant less
than $2^k$ work, cannot generate a message $(FS', \gamma', \beta', P)$ so that processing $(FS', \gamma', \beta', P)$ on
an uncompromised node yields data packet $(FS, \gamma, \beta, P)$. 

To succeed, it is necessary that:
\begin{eqnarray}
	\beta \oplus \{\beta'_{c .. cr - 1}|| 0^c\} = \rho(SV', FS')
	\label{eq:wrap_resistance_1}
\end{eqnarray}
Consider the last $c$ bits of the left side of Equation~\ref{eq:wrap_resistance_1}, we have:
\begin{eqnarray}
	\beta_{[c(r-1)..cr-1]} = \rho(SV', FS')_{[c(r-1)..cr-1]}
	\label{eq:wrap_resistance_2}
\end{eqnarray}
Because $PRG$, $PRP$, $h_{PRG}$, and $h_{PRP}$ are all random oracles, an adversary could generate $FS'$ and $SV'$ that
satisfy Equation~\ref{eq:wrap_resistance_2} only with negligible probability if the adversary performs much less than $2^k$ work.

\subsubsection{Security}
To demonstrate the security property, we need to prove that an adversary with control over all nodes on a path except one node $N$, cannot distinguish among data packets entering $N$. The adversary is able to select paths for the packets traversing $N$ and payloads of the packets. The adversary can also observe packets entering and leaving node $N$ except for packets whose headers match the challenge packets.

We construct the following game $G$. The adversary picks two paths $(n_0, n_1, \dots, n_{\nu-1})$ $0 < \nu \le r$ and $(n_0', n_1', \dots, n_{\nu'-1}')$ $0 \le \nu' \le r$, where $n_i = n_i' \; \forall 0\le i \le j$ and
$n_j = n_j' = N$. Note that the nodes after $N$ in both paths are not necessarily the same set of nodes, and the lengths of the paths can also be different. 
The adversary chooses the public/private key pairs and $SV_i$($SV_i'$) for all nodes except $N$ and can arbitrarily select payload $M$.

The challenger picks randomly a bit $b$ and proceeds in one of the following two ways:

{\bf $b=0$}: The challenger creates an \aheader $(FS_0, \gamma_0, \beta_0)$ through the \name setup phase using 
the path $(n_0, n_1, \dots, n_{\nu-1})$ and uses it to construct a data packet with onion encrypted payload $M^e$ from $M$. 
The challenger outputs $(FS_0, \gamma_0, \beta_0, M^e)$, which could be sent to $n_0$.

{\bf $b=1$}: The challenger creates an \aheader $(FS_0, \gamma_0, \beta_0)$ using the alternative path $(n_0', n_1', \dots, n_{\nu-1}')$ instead and outputs \\$(FS_0, \gamma_0, \beta_0, M^e)$, which could be sent to $n_0'$.

Given the output $(FS_0, \gamma_0, \beta_0)$, the adversary's goal is to determine $b$. 
The adversary can also input any messages $(FS', \gamma', \beta', {M^e}')$ to the honest node $N$ and observes the output messages as long as $(FS', \gamma', \beta') \neq (FS_j, \gamma_j, \beta_j)$.\footnote{We follow the definition of security property~\cite{camenisch2005formal} and only care about header uniqueness.}

We define the adversary's advantage as the difference between $\frac{1}{2}$ and the probability that the adversary succeeds. We will show that the adversary's advantage is negligible. Therefore, the adversary has no better chance to determine $b$ than random guessing.

\paragraph{Proof} (Sketch) We adopt the hybrid-game method. First, we construct a modified game $G_1$ with exactly the same definition, except that we require $j=0$. An adversary who can win $G$ can thus immediately win $G_1$. On the other hand, because the adversary controls nodes $(n_0, \dots, n_{j-1})$ ($(n_0', \dots, n_{j-1}')$) and can thus emulate their processing, the adversary can also win game $G$ if he/she can win game $G_1$. Therefore, the adversary can win game $G$ if and only if the adversary can win game $G_1$.

We create a second game $G_2$, which is the same as $G_1$ except that $FS_0$, $\beta_0$, and $\gamma_0$ are all randomly generated from their corresponding domains. If the adversary can distinguish $G_2$ from $G_1$, we have:
\begin{enumerate}
	\item The adversary can distinguish 
	$$FS_0 = PRP(h_{PRP}(SV_0); R_0||s_0)$$
	from randomness. Then it must be that the adversary is able to tell the output of a pseudo-random permutation with a random key ($h_{PRP}(SV_0)$) from random bits. The probability of success for the adversary is negligible.
	\item The adversary can distinguish 
	$$\beta_0 = PRG(h_{PRG}(SV_0)) \oplus \{FS_1||\gamma_1||\beta_1\}$$
	from randomness. Then it must be the adversary is able to distinguish the output of a secure pseudo-random number generator with a random key ($h_{PRG}(SV_0)$) from randomness. The probability that the adversary succeeds is negligible.
	\item The adversary can distinguish 
	$$\gamma_0 = MAC(h_{MAC}(SV_0); \beta_0)$$
	from randomness. Then it must be the adversary is able to distinguish the output of $MAC$ with a random key $h_{MAC}(SV_0)$ from randomness. Under our random oracle assumption for $MAC$, the probability of success is negligible.
	\end{enumerate}
Therefore, the adversary cannot distinguish $G_2$ from $G_1$. 

Lastly, because in $G_2$, $(FS_0, \gamma_0, \beta_0)$ are all random, the adversary's advantage is 0. Moreover, in our chain of game $G\rightarrow G_1 \rightarrow G_2$, the adversary can only distinguish a game from its previous game with negligible probability. As a result, the adversary's advantage in game $G$ is negligible. \qed

\subsection{Passive De-anonymization}  

\paragraph{Session linkage}
Each session is established independently from every other session, based on fresh, randomly generated keys. Sessions are in particular not related to any long term secret or identifier of the host that creates them. Thus, two sessions from the same host are unlinkable, \ie they are cryptographically indistinguishable from sessions of two different hosts.

\paragraph{Forward/backward flow correlation}
The forward and backward headers are derived from distinct cryptographic keys
and therefore cannot be linked. Only the destination is able to correlate
forward and backward traffic, and could exploit this to discover the round-trip
time (RTT) between the source and itself, which is common to all low-latency
anonymity systems. Sources willing to thwart such RTT-based attacks from
malicious destinations could introduce a response delay for additional
protection.

\paragraph{Packet correlation}
\name obfuscates packets at each hop. This prevents an adversary who observes
packet bit patterns at two points on a path from linking packets between those
two points. In addition to onion encryption, we also enforce this obfuscation
by padding the header and the payload to a fixed length, thwarting packet-size-based
correlation.\footnote{A bandwidth-optimized alternative would be to allow two
or three different payload sizes, at the cost of decreased anonymity.} While this
does not prevent the adversary from discovering that the same flow is passing
his observation points using traffic analysis, it makes this process
non-trivial, and allows upper-layer protocols to take additional measures to
hide traffic patterns.  The hop-by-hop encryption of the payload also hides the
contents of the communication in transit, protecting against information leaked
by upper layer protocols that can be used to correlate packets.

\paragraph{Path length and node position leakage}
\name protects against the leakage of a path's length and of the
nodes' positions on the path (\ie the relative distance, in hops, to the source
and the destination).
In the setup phase, this protection is guaranteed by \sphinx, so only the
common header and FS Payload are subject to leakage (see Section~\ref{sec:packetstructure} for the exact structure of the packets). It is straightforward to
see that the common header does not contain path or position information. The
FS Payload length is padded to the maximum size, and remains constant at each hop
(see Algorithm~\ref{alg:add_fs}). After adding its FS to the front of the FS
Payload, each node re-encrypts the FS payload, making it infeasible for
the next nodes to see how many FSes have previously been inserted.

During data transmission, neither the common header nor
the data payload contain information about path length or node position, so
only the \aheader (anonymous header) needs to be analyzed. The \aheader is
padded to a maximum length with random bytes, and its length remains constant
as it traverses the network (see Algorithm~\ref{alg:get_fs}). The FSes
contained in the \aheader are onion encrypted, as is the padding added at each
hop. Thus, it is not possible to distinguish the initial random padding from
the encrypted FSes, and neither of these from encrypted padding added by the
nodes.

\paragraph{Timing for position identification}
A malicious node could try to learn its position on the path of a session by
measuring timing delays between itself and the source (or the destination) of
that session. 
\name offers two possible countermeasures.  In the first, we assume that the malicious node wishes to measure the network delay between itself and the source.
To perform such a measurement, the node must observe a packet directed to the source (\ie on the backward path) and then observe a response packet from the source (on the forward path).
However, \name can use asymmetric paths~\cite{he2005routing}, making this attack impossible if the single node is not on both forward and backward paths.

The second countermeasure is that, even if the node is on both paths, it is still
non-trivial to discover that a specific forward flow corresponds to a certain
backward flow,
since the forwarding segments for the two paths are independent.
To link the forward and backward flows together the node would need to rely on
the traffic patterns induced by the upper-layer protocols that are running on
top of \name in that session.

\subsection{Active De-anonymization}

\paragraph{Session state modification}
The state of each node is included in an encrypted FS. During the session
setup, the FSes are inserted into the FS payload, which allows the source to
check the integrity of these FSes during the setup phase. During data transmission, FSes are integrity-protected as well through per-hop MACs computed by the source. In this case, 
each MAC protecting an FS is computed using a key
contained in that FS. This construction is secure because every FS is encrypted
using a PRP keyed with a secret value known only to the node that created the
FS: if the FS is modified, the authentication key that the node obtains after
decryption is a new pseudo-random key that the adversary cannot control. Thus,
the probability of the adversary being able to forge a valid MAC is still
negligible.

\paragraph{Path modification}
The two \name data structures that hold paths (\ie FS payloads in the setup
phase and \aheader{}s), use chained per-hop MACs to protect path integrity and
thwart attacks like inserting new nodes, changing the order of nodes, or
splicing two paths. The source can check such chained per-hop MACs to detect
the modifications in the FS payload before using the modified FS payload to
construct \aheader{}s, and similarly intermediate nodes can detect
modifications to \aheader{}s and drop the altered packets. These protections guarantee path information integrity as stated in Section~\ref{sec:desired_properties}.

\paragraph{Replay attacks} Replaying packets can facilitate some types of
confirmation attacks~\cite{Raymond2001}. 
For example, an adversary can replay packets with a
pre-selected pattern and have a colluding node identify those packets
downstream. \name offers replay protection through session expiration; replayed
packets whose sessions have expired are immediately dropped.  Replay of packets
whose sessions are not yet expired is possible, but such malicious behavior can be detected by the end hosts.
Storing counters at the end hosts and including them in the payload ensures that replays are recognizable.
The risk of detection helps deter an adversary
from using replays to conduct mass surveillance. Furthermore, volunteers can monitor the
network, to detect malicious activity and potentially identify which nodes or group of nodes are likely to be misbehaving. 
Honest ASes could control their own nodes as part of an intrusion detection system.

\subsection{Payload Protection}

\paragraph{Payload secrecy}
Data packet payloads are wrapped into one layer of encryption using the key shared between the source and the destination, both for packets sent by the source on the forward and for packets sent by the destination on the backward path (see Section~\ref{sec:DataTransmissionProtocolDescription}).
Assuming that the cryptographic primitives used are secure, the confidentiality of the payload is guaranteed as long as the destination is honest.
In Section~\ref{sec:limitations} we discuss the guarantees for perfect forward secrecy for the data payload.

\paragraph{Payload tagging or tampering}
\name does not use per-hop MACs on the payload of data packets for efficiency
and because the destination would not be able to create such MACs for the packets it sends (since the session keys of the nodes are known only to the source). The lack of
integrity protection allows an adversary to tag payloads.
Admittedly, the use of tagging, especially in conjunction with replay attacks, allows the adversary
to improve the effectiveness of confirmation attacks.
However, end-to-end MACs protect the integrity of
the data, making such attacks (at a large scale) detectable by the end hosts.

\subsection{Denial-of-Service (DoS) Resilience}

\begin{sloppypar}
\paragraph{Computational DoS} The use of asymmetric cryptography in the
setup phase makes \name vulnerable to computational DoS attacks, where
adversaries can attempt to deplete a victim node's computation capability by
initiating a large number of sessions through this node. To mitigate this
attack, \name nodes can require each client that initiates a session to solve a
cryptographic puzzle~\cite{dean2001puzzle} to defend against attackers with limited
computation power. Alternatively, ISPs offering \name as a service can selectively allow
connections from customers paying for the anonymity service.
\end{sloppypar}

\paragraph{State-based DoS} \name is not vulnerable to attacks where adversaries maintain a large number of active sessions through a victim node. One of \name{}'s key features is that all state is carried within packets, thus no per-session memory is required on nodes or rendezvous points.

\subsection{Topology-based Analysis}
\label{sec:ilt}

\begin{figure*}[!htbp]
	\centering
	\subfigure[]{
        \raisebox{9mm}{
		\includegraphics[width=.3\textwidth]{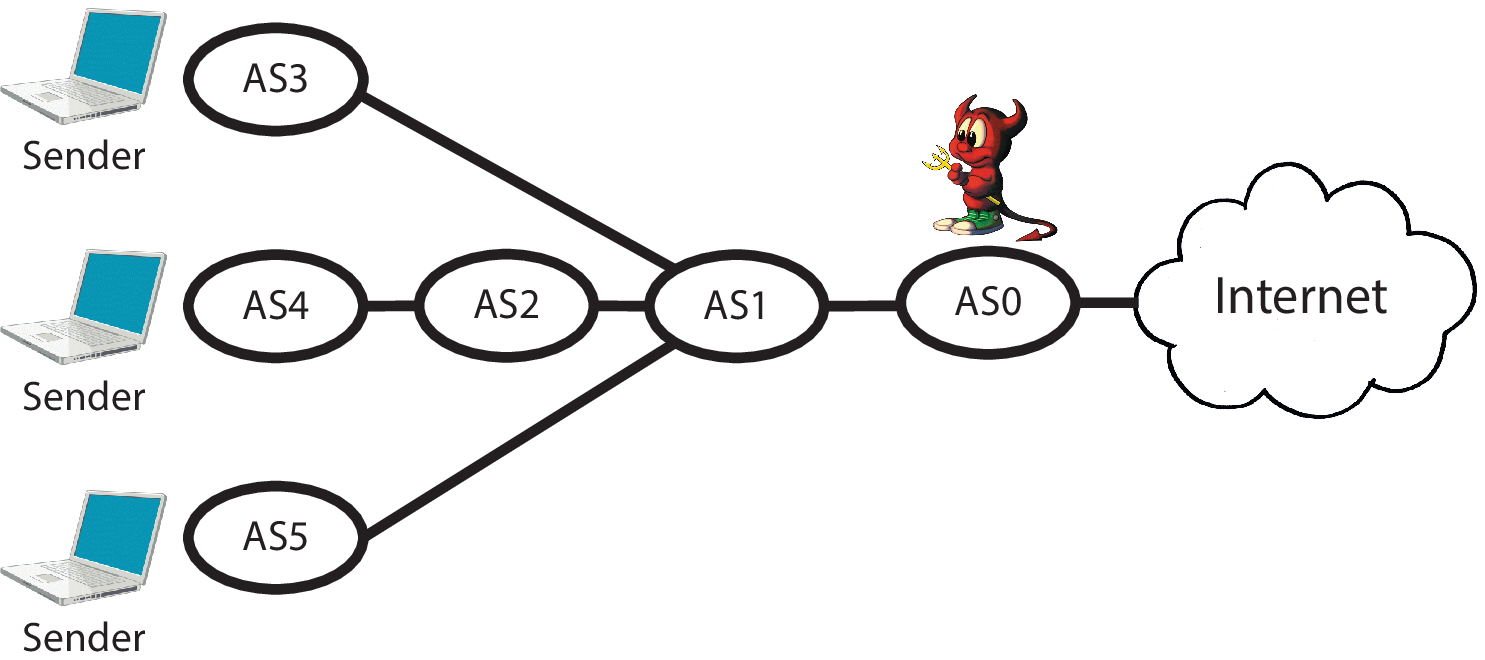}  
    \label{fig:as_position}}
	}
	\subfigure[Without path knowledge]{
		\includegraphics[width=.3\textwidth]{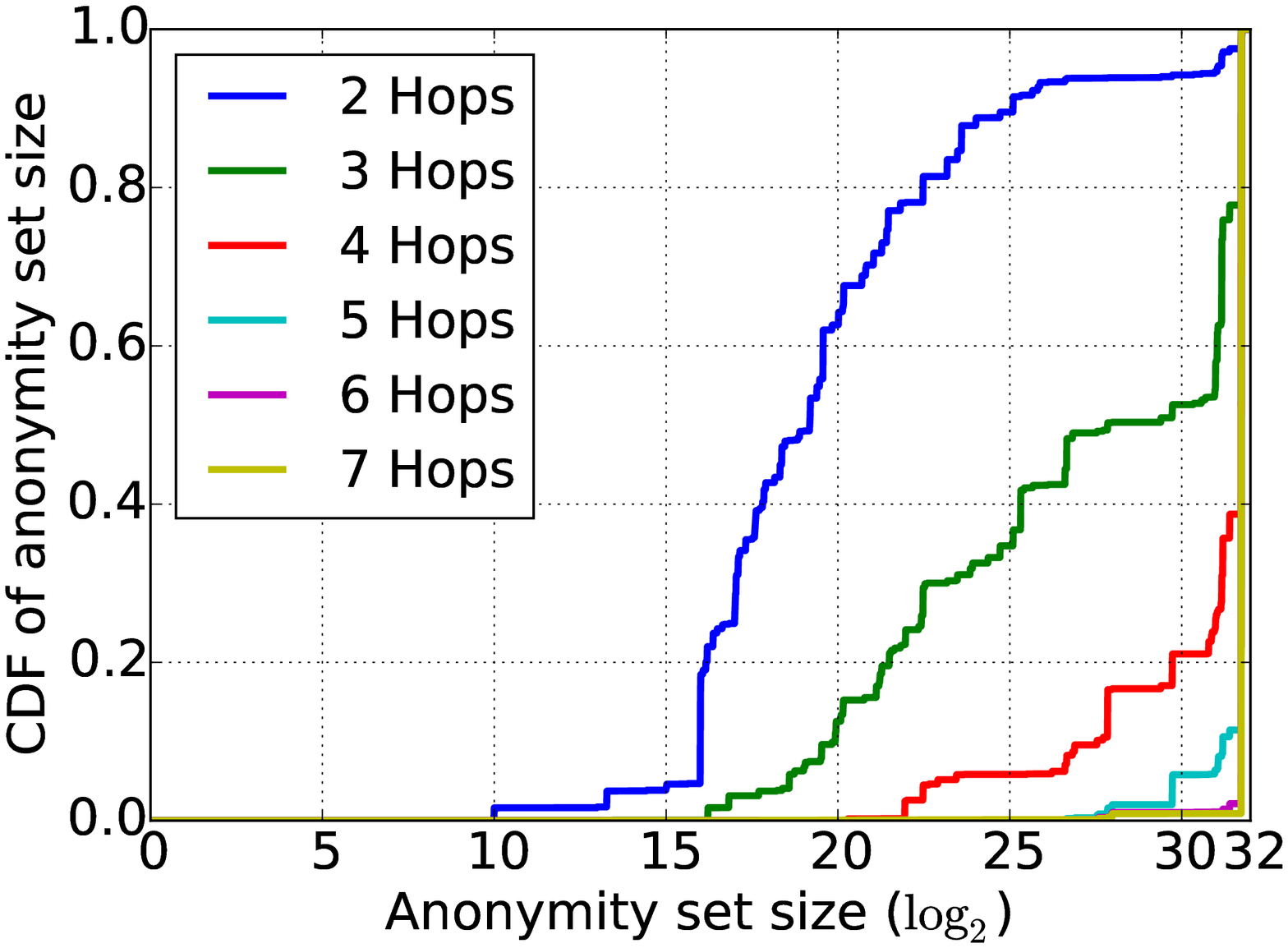}
		\label{fig:anonymityset_subtree_cdf}	
	}
	\subfigure[With path knowledge]{
		\includegraphics[width=.3\textwidth]{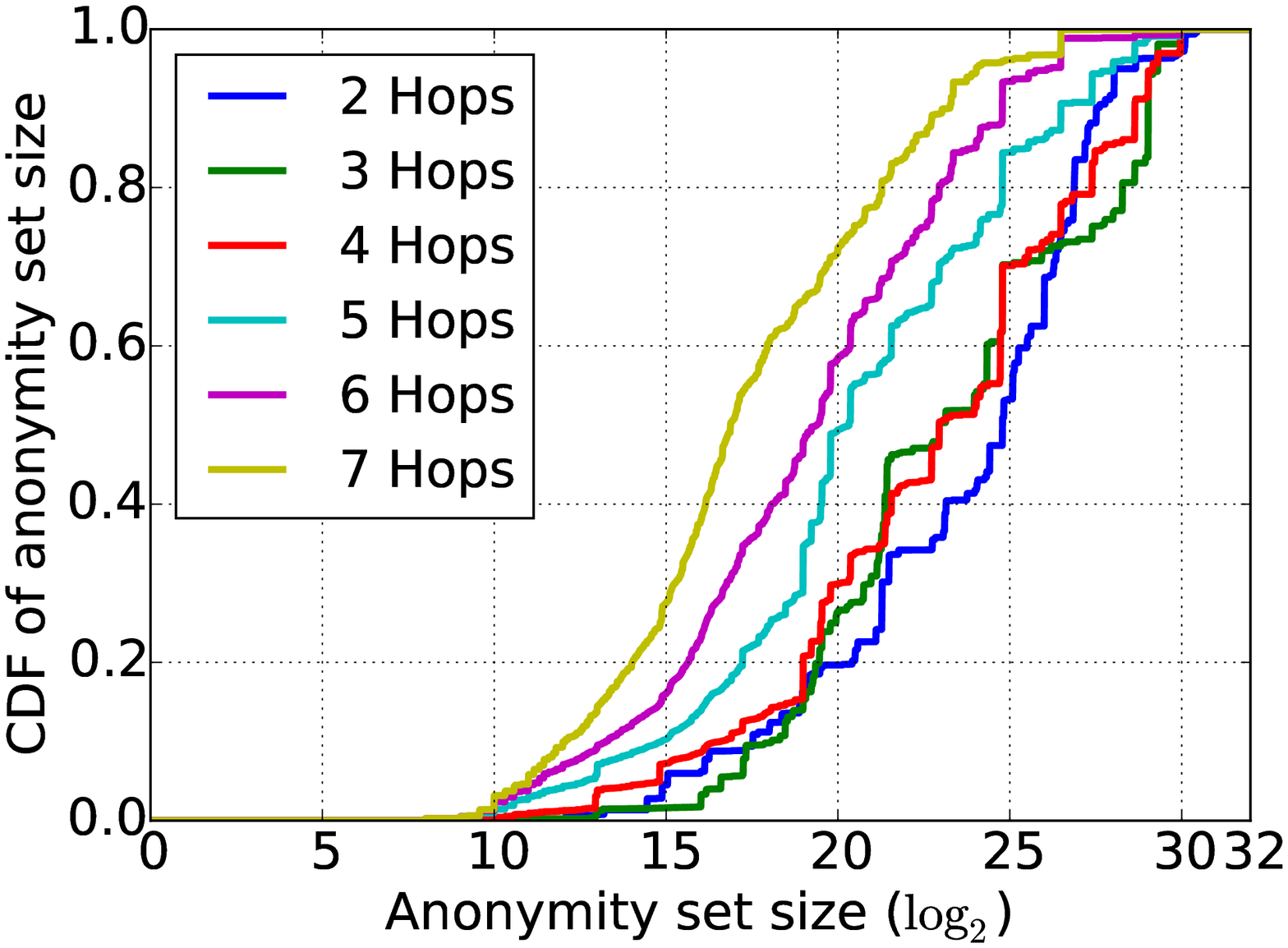} 
		\label{fig:anonymityset_subtree_hops_cdf}
	}
	\caption{a) An example AS-level topology with an adversarial AS (AS0). b) 
	CDF of anonymity-set size when a position-agnostic AS on path is 
	adversarial. ``Hops'' indicates the number of ASes between the adversarial 
	AS and the victim end host. For example, the point (25, 0.4) on the line ``3 hops'' means that the anonymity set size is smaller than $2^{25}$ in 40\% of cases when the end host is 3 hops away from the adversarial AS. c) CDF of anonymity-set size when an adversarial AS knows its own position on the path.
    For Figures b) and c), the maximum size of an end host's anonymity set is $2^{32}$ because we consider the IPv4 address space. Therefore, the ideal case for an end host is that the anonymity set size is $2^{32}$ with probability equal to 1.
    	}

\end{figure*}

Unlike onion routing protocols that use global re-routing through overlay networks (\eg Tor~\cite{dms04} and I2P~\cite{zantout2011i2p}),
\name uses short paths created by the underlying network architecture to reduce latency, and is therefore bound by the network's physical interconnection and ISP relationships. This is an unavoidable constraint for onion routing protocols built into the network layer~\cite{Hsiao2012,Sankey2014}.
Thus, knowledge of the network topology enables an adversary to reduce the number of possible sources (and destinations) of a flow by only looking at the previous (and next) hop of that flow.
For example, in Figure~\ref{fig:as_position}, assume that AS0 is controlled by a passive adversary.
The topology indicates that any packet received from AS1 must have originated from a source located at one of \{AS1, AS2, AS3, AS4, AS5\}. 

We evaluate the information leakage due to the above topology constraints in the scenario where a single AS is compromised.
We derive AS-level paths from iPlane trace-route data~\cite{iplane_dataset},
and use AS-level topology data from CAIDA~\cite{caida_dataset}.
For each AS on each path we assume that the AS is compromised and receives packets from a victim end host through that path. We compute the end host's anonymity set size learned by the adversary according to the topology. For instance, in Figure~\ref{fig:as_position}, if AS0 is compromised and receives from AS1 packets originally sent by a user in AS4, we compute the size of the anonymity set composed of all the ASes that can establish valley-free paths traversing the link from AS1 to AS0. In this example, the anonymity set size would be the sum of the sizes of AS1, AS2, AS3, AS4, and AS5.

Similar to Hsiao \etal~\cite{Hsiao2012}, we use the number of IPv4 addresses to estimate the size of each AS.
Figure~\ref{fig:anonymityset_subtree_cdf} plots the CDF of the anonymity set size for different distances (in number of AS hops) between the adversary and the victim end host. For adversarial ASes that are 4 hops away, the anonymity set size is larger than $2^{31}$ in 80\% of the cases. Note that the maximum anonymity set size is $2^{32}$ in our analysis, because we consider only IPv4 addresses.

\paragraph{Implications of path knowledge}
Knowledge about the path, including the total length of the path and an adversarial node's position on the path, significantly downgrades the anonymity of end hosts. Considering again Figure~\ref{fig:as_position}, if the adversary controlling AS0 sees a packet incoming from AS1 and knows that it is 4 hops away from the source host, he learns that the source host is in AS4. Compared with the previous case, we see that the anonymity set size is strongly reduced.

We quantify additional information leakage in the same setting as the previous evaluation.
Figure~\ref{fig:anonymityset_subtree_hops_cdf} represents the CDFs of the anonymity set sizes of end hosts according to the distance to the compromised AS.
The anonymity set sizes are below $2^{28}$ in 90\% of the cases when the adversarial ASes are 4 hops away, with an average size of $2^{23}$. This average size decreases to $2^{17}$ for the cases where the adversarial ASes are 7 hops away from the target hosts.

Previous path-based anonymity systems designed for the network layer either
fail to hide knowledge about the path~\cite{Sankey2014} or only partially
obscure the information~\cite{Hsiao2012}. In comparison, \name protects both
the path length and the position of each node on the path, which significantly
increases the anonymity-set size.

\section{Evaluation}
\label{sec:evaluation}

\newcommand{\ILT}{ILT\xspace}
\newcommand{\ntor}{L3 \tor}

\begin{sloppypar}
We implemented the \name router logic in an Intel software router using the 
Data Plane Development Kit (DPDK)~\cite{dpdk}. To our knowledge, no other anonymity protocols have been implemented in a router SDK. 
We also implemented the \name client in Python.
Furthermore, we assembled a custom crypto library based on the Intel AESNI cryptographic library~\cite{aesnilibrary}, the curve25519-donna library~\cite{curve25519donna},
and the PolarSSL libraries~\cite{polarssl}. 
We use IP forwarding in DPDK as our performance baseline.
For comparison, we implemented the data forwarding logic from \sphinx, \lap, 
\dovetail, and \tor using DPDK and our cryptographic library.
\end{sloppypar}

Fairly comparing the performance of anonymity systems at the application layer
with those that operate at the network layer is challenging. To avoid
penalizing \tor with additional propagation delay caused by longer paths and
processing delay from the kernel's network stack, we implemented \tor at the
network layer (as suggested by Liu \etal~\cite{liu2011tor}). \tor's design
requires relay nodes to perform SSL/TLS and transport control.  SSL/TLS between neighboring relays at the
application layer maps to link encryption between neighboring nodes at the
network layer, which we consider orthogonal but complementary to \name (see
Section~\ref{sec:comp_other_protocol}).
Hence, for fair comparison, we implemented the network-layer \tor without
SSL/TLS or transport control logic. Throughout our evaluation we refer to this
implementation of \tor as L3 Tor.

\begin{sloppypar}
Our testbed contains an Intel software router connected
to a Spirent TestCenter packet generator and analyzer~\cite{spirent}.
The software router runs DPDK 1.7.1 and is equipped with an Intel Xeon E5-2680 processor (2.70 GHz, 2 sockets, 
16 logical cores/socket), 64 GB DRAM, and 3 Intel 82599ES 40 Gb/s network cards (each with 4 10 Gb/s ports).
We configured DPDK to use 2 receiving queues for each port with 1 adjacent 
logical core per queue. 
\end{sloppypar}

\subsection{Data Forwarding Performance}
\begin{figure}[!htbp]
	\centering
	\includegraphics[width=0.45\textwidth]{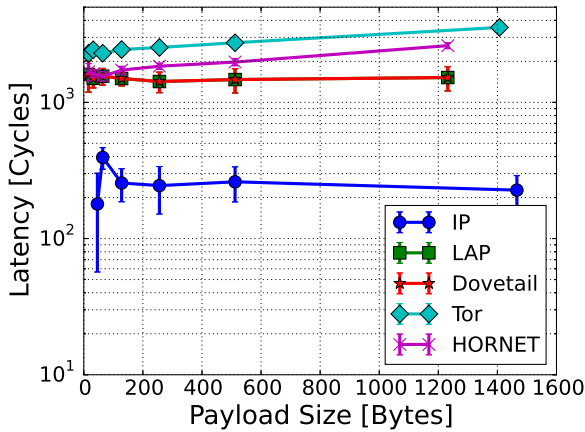}
	\caption{Per-node data forwarding latency on a 10 Gbps link. Lower is better.}
	\label{fig:latency}
\end{figure}

\paragraph{Forwarding latency}
We 
measure the CPU cycles consumed to 
forward a data packet in all schemes. 
Figure~\ref{fig:latency} shows the average latency (with error bars)
to process and forward a single data packet in all schemes (except \sphinx\footnote{We omit Sphinx from the comparison for better readability. In our experiments, processing a \sphinx packet takes more than 640K cycles due to asymmetric cryptographic operations. This is 3 orders of magnitude slower than that of \name, \ntor, \lap, and \dovetail. 
}) when payload sizes vary. 
We observe that
\name, even with onion encryption/decryption over the entire payload and 
extensive header manipulation, is only 5\% slower than \lap and \dovetail for 
small payloads (64 bytes). For large payloads (1200 bytes\footnote{Because \lap, 
\dovetail, and \name all have large packet headers of 300+ bytes, we limit the 
largest payload in our experiments to be 1200 bytes.}), \name is
71\% slower (about 400 nanoseconds slower per packet when using a single core) 
than \lap and 
\dovetail. However, the additional processing overhead enables stronger 
security guarantees.

\paragraph{Header overhead}
\begin{table}
	\centering
	\begin{tabular}{c|ccc}
		\hline
		Scheme & Header Length & Sample Length (Bytes)\\\hline
		\lap & $12 + 2s\cdot r$ & 236 \\
		\dovetail & $12 + s\cdot r$ 
		& 124 \\
		\sphinx & $32 + (2r + 2)s$ & 296 \\
		\tor & $3 + 11 \cdot r$ & 80 \\
		\name & $ 8 + 3r\cdot s$ & 344 \\\hline
	\end{tabular}
	\caption{Comparison between the length of different packet header formats 
	in bytes. $s$ is the length
		of symmetric elements and $r$ is the maximum AS path length. For 
		the sample length, we select $s=16$~Bytes and $r=7$. Analysis 
		of iPlane paths shows that more than 99\% of all paths have fewer than $7$ 
		AS hops.}\label{tab:packet_header_size}
\end{table}
As a result of carrying anonymous session state (specifically cryptographic 
keys) within packet headers, \name  
headers are larger than \sphinx, \ntor, \lap, and \dovetail headers (see 
Table~\ref{tab:packet_header_size}). While larger headers reduce net 
throughput (\ie goodput), this tradeoff appears acceptable: compared to \ntor, 
no state is 
required at relay nodes, enabling scalability; compared to \sphinx, data 
processing speed is higher;  compared to \lap and \dovetail, \name provides 
stronger security properties. 

\paragraph{Goodput}
We further compare all the schemes by goodput, which excludes the header overhead from total throughput. Goodput is a comprehensive metric to evaluate both the packet processing speed
and protocol overhead. For example, a scheme where headers take up a large 
proportion of packets yields
only low goodput. On the other hand, a scheme with low processing speed also results in poor goodput.

\begin{figure*}
  \centering

  \subfigure[40 Byte payloads]{
    \includegraphics[width=0.45\textwidth]{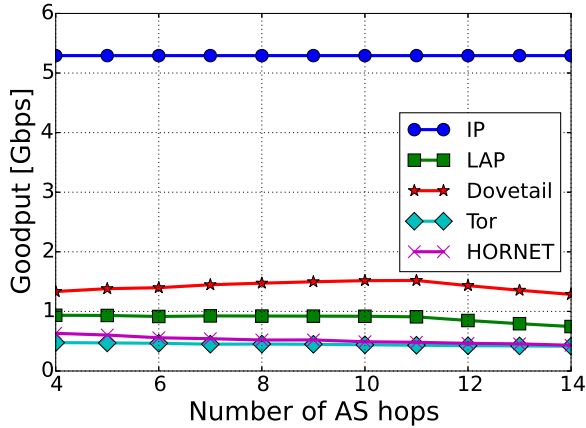}    
    \label{fig:goodput_small}
  }
  \subfigure[1024 Byte payloads]{
    \includegraphics[width=0.45\textwidth]{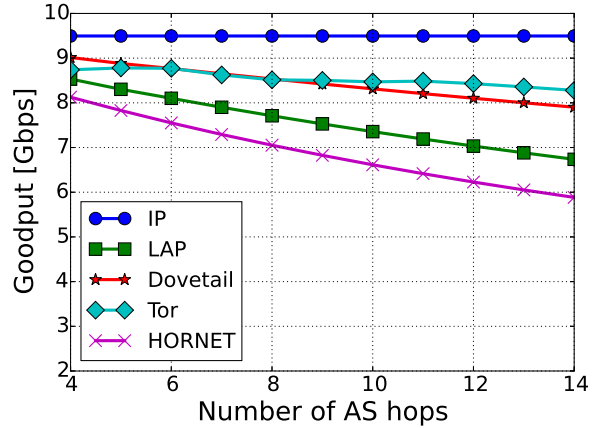} 
    \label{fig:goodput_large}
  }
  \caption{a) Data forwarding goodput on a 10 Gbps link for small packets (40 Byte payloads); b) Data forwarding goodput large packets (1024 Byte payloads). Higher is better.
  	  	}

  \label{fig:goodput}
\end{figure*}

\begin{sloppypar}
Figure~\ref{fig:goodput_small} and Figure~\ref{fig:goodput_large} demonstrate
the goodput of all schemes (except \sphinx\footnote{\sphinx{}'s goodput is less
than 10 Mb/s in both cases because of its large packet headers and 
asymmetric cryptography for packet processing.}) on a 10 Gb/s link when
varying the number of hops $r$, with 40-byte and 1024-byte payloads, respectively. 
Larger $r$ means larger header sizes, which reduces the resulting goodput.
\end{sloppypar}

\smallskip
When the payload size is small, the goodput of all protocols remains stable. 
This is due to the fact that no scheme can saturate
the link, and accordingly the goodput differences between the three schemes 
mainly reflect the different processing latencies among them.
Consequently, \ntor's and \name{}'s goodput is 32\% less than that of \lap and 
\dovetail. 
On the other hand, when the payload size is large, all schemes except \sphinx 
can saturate the 10 Gb/s link. \name can reach 87\% of \lap's 
goodput while providing stronger security guarantees.

\vspace{1em}
\subsection{Max Throughput on a Single Router}
\label{sec:throughput}
To investigate how our implementation scales with respect to the number 
of CPU cores, we use all 12 ports on the software router, generating \name data 
packets at 10 Gb/s on each port. Each packet contains a 7 AS-hop header 
and a payload of 512 bytes, and is distributed uniformly among the working 
ports. We monitor the aggregate throughput on the 
software router.

\smallskip
The maximal aggregate 
throughput of HORNET forwarding in our software router is 93.5 Gb/s, which is comparable to today's switching 
capacity of a commercial edge router~\cite{asr1000}.
When the number of cores ranges from 1 to 4, our \name implementation can 
achieve full line rate (\ie 10 Gb/s per port). As the number of cores  
increases to 5 and above, each additional port adds an extra 6.8Gb/s.

\vspace{1em}
\subsection{Session Setup Performance}
We evaluate the latency introduced by processing setup packets 
 on each border router.
Similar to measuring the latency of data forwarding, we also instrument the 
code to measure CPU cycles consumed to process packets in the session setup 
phase.
Table~\ref{tab:latency_conn_init} lists the average per-node latency for 
processing the two setup 
packets in \name{}'s session setup phase. Due to a Diffie-Hellman key exchange, 
processing the two setup packets in the session setup phase increases 
processing latency (by about 240$\mu$s) compared to data packet processing. 
However, \name must only incur this latency once per session. 

\begin{table}[!htbp]
  \centering
  \begin{tabular}{c|c|c}
    \hline
    Packet & Latency (K cycles) & Latency ($\mu$s) \\
    \hline
    \setupfirst & 661.95 $\pm$ 30.35 & 245.17 $\pm$ 11.24 \\
    \setupsecond & 655.85 $\pm$ 34.03 & 242.91 $\pm$ 12.60 \\\hline
  \end{tabular}
  \vspace{1em}
  \caption{Per-node latency to process session setup packets with standard errors.}
  \label{tab:latency_conn_init}
\end{table}

\subsection{Network Evaluation}
\label{sec:network_evaluation}

\begin{figure}[!htbp]
	\centering
	\includegraphics[width=8cm]{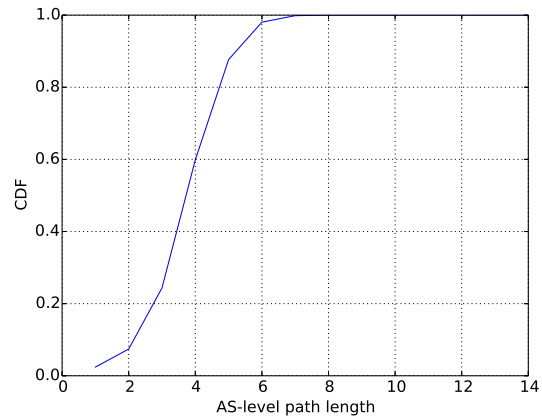}
	\caption{CDF of AS-level path length.}
	\label{fig:as_path_len}
\end{figure}

\paragraph{Distribution of AS-level path length} The bandwidth overhead of a \name packet depends on the number of ASes traversed by the packet. Figure~\ref{fig:as_path_len} demonstrates the CDF of 
AS-level path lengths of the paths extracted from our data source. We observe that 99\% of the paths
have a path length smaller than 7, and the mean AS-level path length is 4.2. Thus, to achieve 128 bits of security, 48 bytes per AS hop are required, leading to an average overhead of 201.6 bytes.

\paragraph{Non-scalability of a stateful design}

We evaluate the memory capacity needed to maintain state required by a stateful design to support Internet-scale anonymous communication. We consider the design of Tor, one of the most popular onion routing systems today~\cite{dms04}, and assume that each Tor node (\emph{onion router} or OR) would correspond to an autonomous system (AS), as proposed by Liu \etal~\cite{liu2011tor}. Analyzing the CAIDA Internet Traces~\cite{caida-passive2014}, we found that a 10 GbE backbone link handles about 1M new flows every minute under normal operating conditions. Since the largest inter-AS links today have up to ten times that capacity (100 Gbps)\footnote{E.g., see \url{www.seattleix.net/participants.htm}.}, this means that at the core of the network there are edge routers of ASes that handle about 10M new flows per minute.

If we assume that half of these flows would use a Tor circuit, because of the default lifetime of circuits of 10 minutes\footnote{We measure the number of flows taking this lifetime into account, in particular we expire flows only if no packets where seen on them for over 10 minutes. Also note that in our setting it would not be possible to have multiple streams per circuit, unless the destinations of those streams are all within the same AS.} we obtain that ORs on such edge routers would need to store state for approximatively 50M circuits at any given time. Since Tor stores at least 376 bytes per circuit, this translates to almost 20 GB of memory. This might still be acceptable for high-end devices, but there are a number of additional factors that make keeping state unfeasible, even for ASes handling less traffic:
\begin{itemize}
	\item The growing number of users on the Internet and the increasing number of devices per user result in an increasing number of traffic flows;
	\item The state for each circuit would actually be larger, as for active circuits the ORs need to store the packets being transmitted until they are acknowledged by the next hop;
	\item A DDoS attack could force an OR to store much more state by opening a large number of new circuits through that OR.
\end{itemize}

\vspace{1em}
\section{Discussion}
\label{sec:discussion}

\subsection{Retrieving Paths Anonymously in FIAs}
\label{sec:discussion:retrieve_path}
\name assumes that the source can obtain a forward path and a backward path to an intended destination  anonymously in FIAs. We briefly discuss how a source host using \name can retrieve two such paths in \nira, \scion and Pathlets. 

\scion hosts rely on path servers to retrieve paths.  In \scion, each
destination node registers on a central server its ``half'' path: the path
to/from the network ``core''. To compose full paths (forward and backward
paths) between a source and a destination, the source only needs to anonymously
fetch the destination's half paths from/to the network core and combine them 
with its own half paths. 

To anonymously retrieve a destination's half paths, the source can use one of the following two methods.
As a first method,
the source can obtain the path to/from a path server through an unprotected query using other schemes, from resolver configuration, or from local services similar to DHCP. The source then establishes an anonymous \name session to the server. Once a \name session is created, the source can proceed to anonymously request half paths of the destination.
Though it is possible to reuse the established \name session to a path server to query multiple paths (for different destinations) for better efficiency, using a separate session to retrieve each path is more secure because it prevents profiling attacks.

\begin{sloppypar}
Alternatively, the source can leverage a private information retrieval (PIR)
scheme~\cite{Chor1998} to retrieve the path anonymously from the path server, so that the path
server cannot distinguish which destination the source connects to. However, a
PIR scheme will inevitably add bandwidth and computational overhead to both the
source and the path server, increasing session setup phase
latency~\cite{mittal2011pir}.
\end{sloppypar}

In \nira and Pathlets, the situation is different because routing information
(\ie inter-domain addresses and route segments, and pathlets, respectively) is disseminated to users. The source can therefore
keep a database local path database, querying it (locally) on demand.

\subsection{Integrating with Security Mechanisms\\ at Different Layers} 
\label{sec:comp_other_protocol}
At the network layer, \name can benefit from ASes that
offer traffic redirection to mitigate topology-based attacks (see Section~\ref{sec:ilt}). For instance, ASes can allow paths that deviate from the valley-freeness policy to
increase the anonymity set size of end hosts. This enables a trade-off between
path length and anonymity, as described by Sankey and Wright~\cite{Sankey2014}.

In addition, upper-layer anonymity protocols can be used in conjunction with \name to provide stronger anonymity guarantees.
For example, to entirely remove the concerns of topology-based attacks, 
a single-hop proxy or virtual private network (VPN) could be used to increase 
the size of the anonymity sets of end hosts. Similar solutions could also 
protect against upper-layer de-anonymization attacks, in particular 
fingerprinting attacks on the transport protocol~\cite{Smart2000}.

At lower layers, \name is also compatible with link-layer protection such as link-level encryption. The role of 
link-level encryption in \name is comparable to SSL/TLS in Tor. Link encryption prevents an adversary eavesdropping
 on a link from being able to distinguish individual sessions from each other, therefore making confirmation attacks much harder for this type of adversary.

\subsection{Limitations}
\label{sec:limitations}

\paragraph{Targeted confirmation attacks}
When for a certain session an adversary controls both the node closest to the
source and the node closest to the destination (or the destination itself), it
can launch confirmation attacks by analyzing flow dynamics.These attacks can be made more effective by replaying packets.

\name, like other low-latency onion routing schemes~\cite{dms04}, cannot
prevent such confirmation attacks targeting a small number of specific
users~\cite{SS03,DBLP:conf/ccs/JohnsonWJSS13}.
However, \name raises the bar of deploying such attacks at scale:
the adversary must be capable of controlling a significant
percentage of ISPs often residing in multiple geopolitical areas. In
addition, the packet obfuscation measures built into \name (discussed in
Section~\ref{sec:security}) make it non-trivial to link two flows, since it is
not possible to simply match packets through bit patterns. Timing
intervals for packet sequences need to be stored and compared, thus performing such operations
for a large fraction of the observed flows is expensive.
Furthermore, it is difficult for attackers to perform active attacks (e.g.,
packet replay) at scale while remaining undetected. For instance, a downstream
benign AS can detect replayed packets by a compromised upstream AS; end hosts
can also detect and report packet tagging attacks when (a threshold number of) end-to-end MACs do not
successfully verify.

\paragraph{Perfect forward secrecy}
A drawback of \name's efficiency-driven design is that it does not provide
perfect forward secrecy for the link between communicating parties.
This means that an adversary could record the observed traffic (the setup
phases, in particular), and if it later compromises a node, it learns which node was next on the path for
each recorded session.
This is an unavoidable
limitation of having a setup that consists of a single round-trip.

Other systems (e.g., \tor) use a telescopic setup\footnote{In the telescopic
setup, a source iteratively sets up a shared key with each AS: the source sets
up a shared key with the first-hop AS; the source sets up a shared key with the
$n$th-hop AS through the channel through 1st-hop AS to $(n-1)$th-hop AS.},
which achieves perfect forward secrecy at the cost of diminished
performance (in particular higher latency, and also an additional asymmetric cryptographic
operation per node). Using a telescopic setup is also possible for
\name, but in addition to the performance cost it also requires that
all paths be reversible. However, this requirement does not hold in today's Internet, where a
significant fraction of AS-level paths are asymmetric~\cite{he2005routing}. 

It is important to note that in \name it is still possible to achieve perfect
forward secrecy for the contents of the communication, i.e., for the data
exchanged between sources and destinations. The destination needs to
generate an ephemeral Diffie-Hellman key pair, and derive an additional
shared key from it.\footnote{This feature,
though omitted in Section~\ref{sec:protocol} for simplicity, is part of our
implementation. It is done in such a way that the forward secret shared key is
included in the destination's FS during the setup, without any additional
packet being required.} Destinations also need to generate a new local
secret $SV$ frequently, so in the event of a destination being compromised it is not possible
for the adversary to decrypt FSes used in expired sessions.

\section{Related Work}
\label{sec:relatedwork}
\paragraph{Anonymity systems as overlays}
The study of anonymous communication began with Chaum's proposal for mix
networks~\cite{Chaum81}. A number of
message-based mix systems have been proposed and deployed
since~\cite{DBLP:conf/ndss/GulcuT96,mixmaster-spec,DBLP:conf/sp/DanezisDM03,Danezis2009}.
These systems can withstand an active adversary and a large fraction of
compromised relays, but rely on expensive asymmetric primitives, and message
batching and mixing. Thus, they suffer from large computational overhead and
high latency.

\begin{sloppypar}
Onion routing systems~\cite{ReSyGo98,Boucher2000Freedom,
brown2002cebolla,dms04} were proposed to efficiently support interactive
traffic.  In general, low-latency onion routing systems are
vulnerable to end-to-end confirmation attacks~\cite{timing-fc2004}, and may
fail to provide relationship anonymity when two routers on the path are
compromised~\cite{DBLP:conf/ih/GoldschlagRS96,DBLP:conf/ccs/JohnsonWJSS13}. \name shares these limitations.
\end{sloppypar}

One specific onion routing system, Tor, has a number of security advantages over \name. Tor can prevent
replays and has perfect forward secrecy for its sessions. Additionally, due to
its overlay design which uses global redirection, Tor is not constrained by the
underlying network topology. However, global redirection enables the attack vector that allows even single compromised ASes to perform confirmation attacks~\cite{murdoch-pet2007, bauer2007low}, as one AS can be traversed multiple times. This attack is not possible in \name since packets traverse each AS on the path only once.

In addition, \name{}'s performance also distinguishes it from all existing schemes
based on overlay networks: first, \name can directly use short paths provided
by underlying network architectures, reducing propagation latency; second,
\name requires only a single round trip to establish a session, reducing the
setup delay; third, \name eliminates the processing and queuing delays both on
relay nodes and in the kernel's network stack; finally, edge routers in \name
offer higher throughput compared to voluntarily-contributed end hosts,
increasing the total throughput of anonymous traffic.

\paragraph{Anonymity systems in FIAs} Hsiao \etal~\cite{Hsiao2012} explored the design space of efficient anonymous systems with a relaxed adversary model. In their scheme, \lap, the adversary can compromise only a single node, and the first hop must always be honest.
Sankey and Wright proposed Dovetail~\cite{Sankey2014} (based on Pathlets~\cite{godfrey2009pathlet} and SCION~\cite{Xin2011SCION,scion2015}) which has the same attacker model as LAP, except it allows the first hop to be compromised. Moreover, neither \lap nor \dovetail can support asymmetric paths where packets traverse different sets of nodes in different directions.
\name offers three improvements over \lap and \dovetail: 1) \name fully hides
path information, \ie total path length and nodes' positions, in packet
headers; 2) \name protects and obfuscates packet contents by
onion-encryption/decryption, thwarting correlating packets of the same flow by
selectors; 3) \name supports asymmetric paths and allows the first hop ASes to
be compromised. Though \name introduces additional overhead in comparison with
\lap and \dovetail, our evaluation results show that \name can still support
high-speed packet forwarding at nearly 80\% of line rate.

The research community has also explored applying onion routing to FIAs. 
Liu \etal~\cite{liu2011tor} proposed Tor instead of IP as an FIA
that regards anonymity as the principal requirement for the network architecture. However, details on how to scale Tor's current design (requiring per-circuit state) to Internet scale were not addressed. 

DiBenedetto \etal~\cite{dibenedetto2011andana} proposed ANDaNA,
to enable onion routing in Named Data Networking (NDN)~\cite{zhang2014named}.
NDN focuses on content delivery and thus inherently different from the FIAs we considered.

\section{Conclusion}
\label{sec:conclusion}
In this paper, we address the question of ``what minimal mechanism can we use to
frustrate pervasive surveillance?'' 
and
study the design of a high-speed anonymity system supported by the network
architecture.
We propose \name, a scalable and high-speed onion routing scheme for 
future Internet architectures. \name nodes can process anonymous traffic at 
over 93 Gb/s and require no per-flow state, paving the path for Internet-scale 
anonymity. Our experiments show that small trade-offs in packet header size 
greatly benefit security, while retaining high performance.

\section{Acknowledgments}
We are grateful for
insightful discussions with Ian Goldberg,  Michael Markus, and the members of the ETH Z\"{u}rich
Network Security group for their discussions and feedback.

The research leading to these results received funding from the European
Research Council under the European Union's Seventh Framework Programme
(FP7/2007-2013) / ERC grant agreement 617605. George Danezis is supported 
by the EU H2020 Project PANORAMIX (653497) and EPSRC Project on ``Strengthening 
anonymity in messaging systems'' (EP/M013286/1). We also gratefully acknowledge
support by ETH Z\"{u}rich, and by Intel for their equipment donation that enabled
the high-performance experiments. 

 \balance
\bibliographystyle{plain}
\small{
\bibliography{bib}
}

\end{document}